\documentclass[a4paper,12pt,english]{article}

\usepackage{amsfonts,bm,amssymb,euscript,array,babel,cite}
\usepackage{amsmath,amsthm,graphics}
\usepackage[dvips]{epsfig}
\usepackage{slashed}
\usepackage{amsmath}



\newcommand{\be}{\begin{equation}} \newcommand{\ee}{\end{equation}}
\newcommand{\beq}{\begin{equation}} \newcommand{\eeq}{\end{equation}}
\newcommand{\beqa}{\begin{eqnarray}}
\newcommand{\eeqa}{\end{eqnarray}} \newcommand{\eq}[1]{(\ref{#1})}
\def\nn{\nonumber} \def\bea{\begin{eqnarray}} \def\eea{\end{eqnarray}}
\def\obar{\overline}

\newcommand{\barr}{\begin{array}}
\newcommand{\earr}{\end{array}}

\def\a{\alpha}  \def\b{\beta}
 \def\g{\gamma} \def\G{\Gamma}
 \def\d{\delta} \def\D{\Delta}
 \def\e{\epsilon} 
    
 \def\L{\Lambda}  \def\m{\mu}
\def\n{\nu}  \def\p{\pi}  \def\r{\rho}
\def\s{\sigma} \def\S{\Sigma}  \def\th{\theta}
\def\Th{\Theta}

\def\cA{{\cal A}}      
\def\cH{{\cal H}}  \def\cJ{{\cal J}}   \def\cM{{\cal M}} \def\cN{{\cal N}}
\def\cO{{\cal O}}

\def\R{{\mathbb R}} \def\C{{\mathbb C}} 
 \def\one{\mbox{1 \kern-.59em {\rm l}}}

\def\mg{\mathfrak{g}}

\def\LNC{\Lambda_{\rm NC}}

\def\bit{\begin{itemize}} \def\eit{\end{itemize}} 
\def\Tr{\mbox{Tr}}
\def\tr{\mbox{tr}}

\def\({\left(} \def\){\right)}

 \allowdisplaybreaks[3]

\begin{document}

\renewcommand{\title}[1]{\vspace{10mm}\noindent{\Large{\bf
#1}}\vspace{8mm}} \newcommand{\authors}[1]{\noindent{\large
#1}\vspace{5mm}} \newcommand{\address}[1]{{\itshape #1\vspace{2mm}}}

\begin{titlepage}
\begin{flushright}
UWThPh-2011-21
\end{flushright}

\begin{center}

\title{ \Large Intersecting branes and a standard model realization \\[1ex] in matrix models}

\vskip 3mm

\authors{Athanasios {Chatzistavrakidis${}^{*,}${\footnote{than@th.physik.uni-bonn.de}}},
Harold {Steinacker${}^{\dagger ,}${\footnote{harold.steinacker@univie.ac.at}}},  \\[1ex] and
 George {Zoupanos${}^{\ddagger ,}${\footnote{George.Zoupanos@cern.ch}}}}

\vskip 3mm

\address{${}^*$ {\it Bethe Center for Theoretical Physics and Physikalisches Institut der Universit\"{a}t Bonn
\\
Nussallee 12, D-53115 Bonn, Germany}\\[3ex] 
${}^{\dagger}$ Faculty of Physics, University of Vienna \\ Boltzmanngasse 5, A-1090 Wien, Austria \\[3ex] 
${}^{\ddagger}$ Physics Department, National Technical University,\\ Zografou Campus, GR-15780 Athens, Greece}

\vskip 1.4cm

\textbf{Abstract}

\end{center}
We consider intersecting brane solutions of the type IIB matrix model.
It is shown that fermionic zero-modes  arise on such backgrounds,
localized at the brane intersections.
They lead to chiral fermions in four dimensions under certain conditions. 
Such configurations reproduce many of the welcome features in similar string-theoretic constructions. 
Therefore they can be used to construct semi-realistic particle physics models in the framework of Yang-Mills matrix models. 
In particular, we present a brane configuration which realizes the correct chiral spectrum of the standard model 
in the matrix model. 
Furthermore, the stability of intersecting branes is discussed by analyzing the 1-loop effective action.
It is shown that intersecting branes may form a bound state for certain flux configurations.
The four-dimensional geometry of the branes is generic, and determined by the (emergent) gravity
sector of the matrix model.

\vskip 3mm

\begin{minipage}{14cm}%

\end{minipage}

\end{titlepage}

\tableofcontents


\section{Introduction}

A long-standing problem in modern theoretical physics is to achieve a description of nature at the Planck scale, 
where quantum mechanics and gravity both become important. At this fundamental level, frameworks such as 
string theory and non-commutative geometry have provided numerous ideas and hints. On the other hand, our 
knowledge about the natural world to date resides in effective theories, such as the Standard Model (SM), 
which can be tested in accelerators. However, there is no convincing link up to now bridging 
fundamental and effective theories.   

Matrix Models (MM) offer a framework where both profound conceptual problems as well as questions about low-energy physics 
may be addressed. Indeed, 
the MMs introduced by Banks-Fischler-Shenker-Susskind (BFSS) and 
Ishibashi-Kawai-Kitazawa-Tsuchiya (IKKT) are supposed to provide a non-perturbative definition of M theory and type 
IIB string theory respectively \cite{Banks:1996vh,Ishibashi:1996xs}. The latter MM may be also interpreted as a
 non-perturbative formulation of Supersymmetric Yang-Mills (SYM) theory on non-commutative four-dimensional space, 
realized as the Moyal-Weyl quantum plane $\R^4_\theta$ embedded in ambient $\R^{10}$. However, such 
MMs are much richer than the four-dimensional 
gauge theory; in particular, the geometry is dynamical, and the gauge theory is automatically coupled to 
(``emergent'') gravity.
Since we consider that the gauge theory lives on a brane embedded in $\R^{10}$, its effective geometry is governed (to a large extent) by the embedding of the brane. These geometrical degrees of freedom are dynamical and governed by the matrix model, leading to a gauge theory coupled to  dynamical gravity. This effective or ``emergent'' gravity has been clarified and elaborated recently within the MM point of view \cite{Steinacker:2007dq,Steinacker:2010rh}.

Apart from their fundamental significance, MMs are also useful laboratories for the study of structures 
which could be relevant from a low-energy point of view. Indeed, they generate a plethora of interesting solutions, 
corresponding to strings, $D$-branes and their interactions \cite{Ishibashi:1996xs,Chepelev:1997ug,Fayyazuddin:1997zx,Aoki:1999vr}, as well as to non-commutative/fuzzy spaces \cite{Iso:2001mg,Kimura:2001uk,Kitazawa:2002xj,Azuma:2004zq,Steinacker:2003sd}. Such backgrounds naturally give rise to non-abelian gauge theories. Moreover, intersecting non-commutative $D$-brane solutions and their stability were studied in \cite{Huang:2003sx,Bergman:2001kz,Tseng:2000sz,Kim:2000mp,Aganagic:2000mh,Gross:2000ph}{\footnote{See also \cite{Berenstein:2010bi}.}.       
However, these works do not address any issues related to possible phenomenological applications. 
Some attempts to discuss model building in the MM framework were made in \cite{Aoki:2002jt} (see also \cite{Aoki:2010gv}), 
where an orbifold MM was considered, and more recently in \cite{Grosse:2010zq}. In the latter, the SM particles were accommodated in the MM but it was not clear how a chiral spectrum can be achieved.  

From a string-theoretic point of view it is hard to underestimate the impact of $D$-brane model building in the quest 
for phenomenological applications of string theory. $D$-branes in type I and type II string theories provided from 
the beginning the possibility of a bottom-up approach to the string embedding of the SM 
\cite{Antoniadis:2000ena,Antoniadis:2002qm,Aldazabal:2000sa}. Moreover, the study of intersecting brane configurations, 
where chiral fermions can be localized \cite{Berkooz:1996km}, opened up even more possibilities 
for model building in the context of type II orientifolds 
(see \cite{Ibanez:2001nd,Blumenhagen:2005mu,Blumenhagen:2006ci,Marchesano:2007de} 
and references therein). 

In the present paper we explore the possibility to describe realistic low-energy physics in the framework of MMs. 
Since certain aspects of this task have not been touched upon before, we shall develop the necessary formalism 
as needed. Section 2 contains a brief exposition of the necessary facts about the type IIB MM. 
In particular, the generic solutions of the model are identified and its relation to emergent gravity is discussed. 

In section 3 we present more general solutions of the MM, corresponding to multiple intersecting non-commutative brane backgrounds. We mainly focus on the case of flat branes and study in detail fermions in such backgrounds. Subsequently, explicit cases of brane configurations are considered and the zero-mode structure of the fermions on the brane intersections is determined. In particular, we study the cases of two $D5$ branes, one $D5$ and one $D7$ brane, two $D7$ branes and two $D5$ and one $D7$ brane. It is shown that chiral fermions can indeed be localized at the intersections.

Having proven that chiral fermions can be accommodated in the MM, we proceed 
in section 4 to the description of brane configurations which can support the gauge group and the particle spectrum of the SM. Essentially a single such
 configuration is determined, based on four $D7$ branes appropriately embedded in $\R^{10}$ and carrying appropriate fluxes on their compactified 
six-dimensional intersections. Moreover, we comment on the construction of 
other models, based on configurations of mutually intersecting $D5$ and $D7$ 
branes, which may provide additional features such as right-handed neutrinos.
 Finally, at the end of the section we discuss some further issues related to the bosonic sector of the model, 
the Yukawa couplings and the anomalies.

In section 5 we address the issue of the stability of the above 
configurations. Studying the 1-loop effective action of the MM, we argue that 
for certain flux configurations intersecting branes may form a bound state 
and therefore do not collapse into coinciding branes. Section 6 contains our 
conclusions. Finally, in the appendix A we present in more detail some 
computations related to characters of representations of $SO(10)$, which 
are used in section 5.
Appendix B contains a discussion of a supersymmetry structure 
which arises on the brane intersections.

\section{The IKKT matrix model}


\paragraph{Definition of the model}

The IKKT or IIB matrix model was originally proposed in \cite{Ishibashi:1996xs} as a non-perturbative definition of the 
type IIB superstring theory. It is a zero-dimensional reduced matrix model defined by the action
\be
S  =  -{\L^4\over g^2}Tr({1\over 4}[X_a,X_b][X^a,X^b]
+{1\over 2}\bar{\psi}\Gamma ^a[X_a,\psi ]),
\label{IKKT-action}
\ee
where $X_a, a=0,\dots,9$ are ten hermitian matrices, 
and $\psi$ are sixteen-component Majorana-Weyl spinors of $SO(9,1)$.
Indices are raised and lowered with the invariant tensor $g_{ab} = \eta_{ab}$, or 
possibly $g_{ab} = \delta_{ab}$ in the Euclidean version where $SO(9,1)$ is replaced by $SO(10)$.
The $\Gamma^a$ are generators of the corresponding Clifford algebra. 
$\L$ is an energy scale, which we will set 
equal to one $\L=1$, and work with dimensionless quantities. Finally, $g$ is a parameter which can be related to the 
gauge coupling constant.

The symmetry group of the above model contains the $U(N)$ gauge group
(where the limit $N \to \infty$ is understood) as well as the  $SO(10)$ or $SO(9,1)$ global symmetry.
Moreover, the model enjoys a ${\cal N}=2$ space-time supersymmetry (SUSY), realized by the following transformations, 
\bea \d_{\e}^{(1)}\psi&=&\frac i2[X_a,X_b]\G^{ab}\e, \\
		\d_{\e}^{(1)}X_a&=&i\bar\e\G_a\psi, \\
			\d_{\xi}^{(2)}\psi&=&\xi, \\
				\d_{\xi}^{(2)}X_a&=&0. 
\label{susy}
\eea 
Therefore, the amount of SUSY indeed matches that of the type IIB superstring. 
Let us also note that the homogeneous $\e$-supersymmetry is inherited by the maximal ${\cal N}=1$ SUSY of 
Super-Yang-Mills (SYM) theory in ten dimensions.

It is important to stress that due to its 0-dimensional nature, the IKKT model is not defined on any predetermined space-time background. 
Instead, space-time emerges as a particular solution of the model, as we discuss in the following. This picture provides a dynamical origin for geometry and space-time.

\paragraph{Equations of motion and basic solutions}

Varying the action (\ref{IKKT-action}) with respect to the matrices $X_a$ and setting $\psi=0$, the following equations of motion are obtained,
\be\label{ikkteoms} [X_b,[X^a,X^b]]=0. \ee 
Simple as they may appear, these equations admit diverse interesting and non-trivial solutions. 

Clearly, the simplest solution is given by a set of commuting matrices, $[X^a,X^b]=0$. In that case, 
the matrices $X^a$ can be simultaneously diagonalized and therefore they may be expressed as 
\be X^a = \mbox{diag}(X^a_1,X^a_2,\dots,X^a_N). \ee 
However, such solutions are in a sense degenerate and do not lead to interesting dynamics. 

For notational convenience let us now split the ten matrices $X^a$ in two sets; we shall use the following notation,
\be\label{split} X^a=  \begin{pmatrix}
	  X^\mu \\
           Y^{i}
\end{pmatrix}, \ee 
where the $X^{\m},\m=0,\dots,3$ correspond to the first four $X^a$ matrices and the $Y^{i}, i=1,\dots,6$ to the six rest of the 
$X^a$ matrices respectively. 
Let us stress that although a splitting of the type $10=4+6$ is considered here, 
this is not a priori favoured by the matrix model action\footnote{However, there should be some dynamical reason  
within the matrix model why four dimensions are preferred, cf. \cite{Nishimura:2001sx,Kawai:2002ub}.}. 
In this notation, another solution of the equations (\ref{ikkteoms}) is given 
by
\be
 X^a = \begin{pmatrix}
	 \bar X^\mu \\
          0
\end{pmatrix} \ee 
where $\bar X^{\m}$ are the generators of the Moyal-Weyl quantum plane $\R^4_{\theta}$, which satisfy the commutation relation
\be 
[\bar X^\mu,\bar X^\nu] = i \theta^{\mu\nu},
\ee
where $\theta^{\m\n}$ is a constant antisymmetric tensor. This solution corresponds to a single non-commutative (NC) flat 3-brane, 
which corresponds to space-time emerging as a solution of the matrix model. 
Being a single brane, this solution is associated to an abelian gauge theory. An obvious generalization of the above solution is given by 
\be
 X^a = \begin{pmatrix}
	 \bar X^\mu \\
          0
\end{pmatrix}\otimes \one_{n},
\label{coinciding-branes}
\ee 
which is interpreted as $n$ coincident branes carrying a non-abelian $U(n)$ gauge theory.

In the following paragraph we shall briefly argue that deformations of the above
brane, interpreted as more general (curved) submanifolds $\cM^4 \subset \R^{10}$,
are equipped with an effective metric. Therefore the effective field theory living on the
brane is coupled to an effective gravity.

\paragraph{Emergent geometry and gravity.}

The flat solutions $\R^4_\theta$ are special cases of  NC branes with
generic embedding in the ambient $\R^{10}$. Such generic branes are desribed by 
 quantized embedding functions
\be
X^a \sim x^a: \quad \mathcal{M}^{2n}\hookrightarrow \R^{10}
\ee
of a $2n$ dimensional submanifold, where the matrices $X^a \sim x^a$ are interpreted as
quantized embedding functions. Furthermore,
\begin{align}
[X^\mu,X^\nu] \sim  i \{x^\mu,x^\nu\} = i \theta^{\mu\nu}(x)\,
\end{align}
is interpreted as a quantized Poisson structure on $\mathcal{M}^{2n}$.
Here $\sim$ denotes the
semi-classical limit where commutators are replaced by Poisson brackets, and
$x^\mu$ are locally independent coordinate functions chosen among the $x^a$.
Thus we are considering quantized embedded
Poisson manifolds $(\cM^{2n},\theta^{\mu\nu})$.
The sub-manifold $\cM^{2n}\subset\R^{10}$
is equipped with a non-trivial induced metric
\begin{align}
g_{\mu\nu}(x)=\partial_\mu x^a \partial_\nu x^b\eta_{ab}\,,
\label{eq:def-induced-metric}
\end{align}
via the pull-back of $\eta_{ab}$.
It is then not hard to see  \cite{Steinacker:2008ri}
that the kinetic term for
all (scalar, gauge and fermionic) fields in the MM on such a background $\cM^{2n}$
is governed (up to possible conformal factors) by the effective metric
\begin{align}
G^{\m\n}& \sim \th^{\m\r}\th^{\n\s}g_{\r\s}\,,
\label{eff-metric}
\end{align}
so that $G_{\mu\nu}$ must be interpreted as gravitational metric.
Since the embedding is dynamical, the model describes a dynamical
theory of gravity, realized on dynamically
determined submanifolds of $\R^{10}$.

\section{Intersecting branes}

\subsection{Multiple brane backgrounds}

In the previous section we presented the basic solutions of the IKKT matrix model 
corresponding to the four-dimensional Moyal-Weyl quantum plane, or
to $n$ coinciding flat non-commutative 3-branes. 
An obvious generalization is to consider $2n$-dimensional quantum plane solutions of the model,
\be
[X^a,X^b] = i \Theta^{ab},
\label{basic-brane-2n}
\ee
where $\Theta^{ab}$ has rank 2n and is embedded along $\R^{2n}$ in some given subspace.
To emphasize the analogy with string theory,
we will call such a solution a $D(2n-1)$-brane, although there are no strings in the model.
More solutions can be obtained by combining two or more of such solutions, 
embedded via block-matrices
\be
X^a = \begin{pmatrix}
       X^a_{(1)} & 0 \\ 0 & X^a_{(2)}
      \end{pmatrix}
\label{2-branes}
\ee
and similarly with more than two blocks. Furthermore, each block may be replaced by a stack of coinciding branes 
as in \eq{coinciding-branes}.
In the present paper, we will
assume that all these blocks share some common $\R^4_{0123}$ with generators 
\be
X^\mu = \{X^0,X^1,X^2,X^3\} ,
\ee
and contain in addition some extra-dimensional $\R^{2n-4}$ in the transverse $\R^6$,
parametrized in terms of $Y^i, \, i = 1,..,6$. Explicitly\footnote{We hope that the size of the matrix $X^{\mu}$ is clear from the context and the notation below does not cause any confusion.},
\be
X^\mu =\begin{pmatrix}
       X_{(1)}^\mu & 0 \\ 0 & X_{(2)}^\mu
      \end{pmatrix}= \begin{pmatrix}
       \bar X^\mu & 0 \\ 0 & \bar X^\mu
      \end{pmatrix}, \qquad 
Y^i = \begin{pmatrix}
       Y^i_{(1)} & 0 \\ 0 & Y^i_{(2)}
      \end{pmatrix}.
\ee
Now let us consider fermions in such a background, specifically bi-fundamental ones
connecting the branes. We are interested in the zero-modes which are localized at the intersection,
realized as off-diagonal matrices
\be
\Psi = \begin{pmatrix}
       0 & \Psi_{(12)} \\ \Psi_{(21)} & 0 
      \end{pmatrix} \label{Psi}.
\ee
Those are candidates for chiral fermions. In order to study them we need to determine the action of the Dirac operator on them. 
Let us first note that according to the splitting (\ref{split}), the ten-dimensional Clifford algebra, generated by
$\Gamma_a$, naturally separates into
a four-dimensional and a six-dimensional one as follows,
\bea
\Gamma_a&=&(\Gamma_{\mu},\Gamma_i), \nn\\ \Gamma_{\mu}&=&\gamma_{\mu}\otimes\one_8, \nn\\ \Gamma_{i}& =& \gamma_5\otimes\Delta_i.
\eea
Here the $\gamma_{\mu}$ define the four-dimensional Clifford algebra, $\g_5$ is the usual chirality operator in 4D, 
while the $\Delta_i$ define the six-dimensional Euclidean Clifford
algebra. The explicit form of these representations will not be needed in the following and therefore it will 
not be given here\footnote{The interested reader may consult \cite{Chatzistavrakidis:2009ix}.}.
 
 The Dirac operator may be split as
\be\label{Dirac10D}
\slashed{D} \Psi = \slashed{D}_4 \Psi + \slashed{D}_6 \Psi 
 = \slashed{D}_4 \Psi + \Delta_i[Y^i,\Psi],
\ee where $\slashed{D}_4$ is the Dirac operator on the 4D quantum plane and $\slashed{D}_6$ is its internal part, 
\be \slashed{D}_6\cdot=\D_i[Y^i,\cdot]. \ee
Focusing on the internal part $\slashed{D}_6$, which
will lead to localized zero modes at $y^i \approx 0$,
some intuition may be gained by writing
\bea
\slashed{D}_6 \Psi_{(12)} &=& 
\Delta_i[Y^i,\Psi_{(12)}] = \Delta_i (Y^i_{(1)} \Psi_{(12)} - \Psi_{(12)} Y^i_{(2)})   \nn\\
&=& \Delta_i \Big(([Y^i_{(1)}, \Psi_{(12)}] + \Psi_{(12)} Y^i_{(1)})
  + ([Y^i_{(2)}, \Psi_{(12)}] - Y^i_{(2)}\Psi_{(12)} ) \Big) .
\eea
The expression in the second line is reminiscent of the coupling of a fermion to a magnetic field. 
Furthermore, let us consider the square of the internal part of the Dirac operator acting on the mode $\Psi_{(12)}$,
\be
(\slashed{D}_6)^2 \Psi_{(12)} = \Box_6\Psi_{(12)}  + \Sigma_{ij}[\Theta^{ij},\Psi_{(12)} ], \ee
where the Laplacian is defined as
\be \Box_6\cdot = [Y_i,[Y_i,\cdot]], \ee and 
\be \S_{ij}=\frac i4 [\D_i,\D_j]. \ee
Then, noting that
\bea
\Box_6\Psi &=&  (Y^i Y^i) \Psi + \Psi  (Y^i Y^i) - 2 Y^i\Psi Y^i,  \nn\\
\Box_6\Psi_{(12)} &=&   Y^i_{(1)}Y^i_{(1)} \Psi_{(12)} + \Psi_{(12)} Y^i_{(2)}Y^i_{(2)} - 2 Y^i_{(1)} \Psi_{(12)} Y^i_{(2)},
\label{Box-off-diag}
\eea
it is straightforward to show that
\bea
(\slashed{D}_6)^2 \Psi_{(12)} 
&=& Y^i_{(1)}Y^i_{(1)}\Psi_{(12)} + \Psi_{(12)} Y^i_{(2)}Y^i_{(2)}  
  - 2 Y^i_{(1)} \Psi _{(12)}  Y^i_{(2)} \nn\\
&& + \Sigma_{ij}^{(1)}\Theta^{ij}_{(1)}\Psi_{(12)} - \Sigma_{ij}^{(2)} \Psi_{(12)} \Theta^{ij}_{(2)}. 
\eea
For simplicity we focus on the case of branes intersecting with angle $\frac{\pi}2$.
We can then separate the indices $i=1,...,,6$ into subsets where either $Y^i_{(1)}$ or $Y^i_{(2)}$
vanishes. Then $Y^i_{(1)} \Psi _{(12)}  Y^i_{(2)} =0$, and we find
\bea
(\slashed{D}_6)^2 \Psi_{(12)} &=&  Y^i_{(1)}Y^i_{(1)}\Psi_{(12)} + \Psi_{(12)} Y^i_{(2)}Y^i_{(2)} 
 + \Sigma_{ij}^{(1)}\Theta^{ij}_{(1)}\Psi_{(12)} - \Sigma_{ij}^{(2)} \Psi_{(12)} \Theta^{ij}_{(2)} .
\eea
Here $\Theta^{ij}_{(1)/(2)}$ is the non-commutative flux on the brane described by $Y^i_{(1)}$ and $Y^i_{(2)}$ respectively, which using the above assumptions
arises in the perpendicular blocks.

The above analysis corresponds to configurations with two quantum planes. It is straightforward to consider configurations with three quantum planes as well. 
Such configurations are embedded via the following block matrices,
\be
X^a = \begin{pmatrix}
       X^a_{(1)} & 0 & 0 \\ 0 & X^a_{(2)} & 0 \\ 0 & 0 & X^a_{(3)}
      \end{pmatrix}. 
\label{3-branes}
\ee 
Assuming again that they all correspond to space-time filling branes, the matrices may be split in two parts,
\be
X^{\mu} = \begin{pmatrix}
       \bar X^{\mu} & 0 & 0 \\ 0 & \bar X^{\mu} & 0 \\ 0 & 0 & \bar X^{\mu}
      \end{pmatrix}, \qquad Y^i= \begin{pmatrix}
       Y^i_{(1)} & 0 & 0 \\ 0 & Y^i_{(2)} & 0 \\ 0 & 0 & Y^i_{(3)}
      \end{pmatrix}.\ee
The bi-fundamental fermions in this background may be written as
\be \Psi=\begin{pmatrix} 0 & \Psi_{(12)} & \Psi_{(13)} \\ \Psi_{(21)} & 0 & \Psi_{(23)} \\
          \Psi_{(31)} & \Psi_{(32)} & 0
         \end{pmatrix}.
\ee
The Dirac operator is given by (\ref{Dirac10D}), while its internal part on each component is 
\bea 
\slashed{D}_6\Psi_{(ab)}&=&\Delta_i(Y^i_{(a)}\Psi_{(ab)}-\Psi_{(ab)}Y^i_{(b)}),  
\eea
for $a,b = 1,2,3$.
One can then easily write down the formulae for $(\slashed{D}_6)^2$ and $\Box_6$. Later in this paper, in the discussion of the possibility to construct realistic models, 
we shall also consider configurations with four and five quantum planes. By now their treatment should be obvious
and there is no need to keep cluttering with formulae.

Finally, it is important to understand the degrees of freedom in the block matrices.
The diagonal block-matrices are elements in the matrix algebras $End(\cH_a)$ and $End(\cH_b)$ respectively,
interpreted as functions on the corresponding quantum planes 
$\R^{2n}_a$ and $\R^{2n}_b$ represented on their corresponding
Hilbert space ${\cal H}_{a/b}$. In contrast, the
off-diagonal block matrices, such as $\Psi_{(ab)}$ and $\Psi_{(ba)}$, are not elements of an algebra but bi-modules over 
these two algebras acting from the left and from the right respectively. In other words,
\be
 \Psi_{(ab)} \in \cH_a \otimes \cH_b^*, \qquad  
 \Psi_{(ba)} \in \cH_b \otimes \cH_a^* .
\label{block-algebra}
\ee
They can be interpreted as 
oriented modes (''strings``) connecting the branes $a$ and $b$.

\subsection{Explicit cases and chiral fermions}

In the previous subsection we studied general backgrounds of the IKKT matrix model, corresponding to flat intersecting non-commutative branes, and discussed fermions in such backgrounds. 
Let us now proceed to specific cases of brane intersections and study the chirality of the fermions in each case. 

\subsubsection{$\R^2 \cap \R^2$}
\label{sec:D5D5}

In order to understand the basic mechanism, it is instructive to start with the system of two 
$D5$ branes along $\R^2_{45}$ and  $\R^2_{67}$ respectively, sharing a common $\R^4$. 
We will ignore the $\R^4$ part from now on, which is common to all branes under consideration.
Then we can write
\bea
Y^i_{(a)} = \left\{\begin{array}{c}\, y^i, \quad i \in\{1,2\} \\ 0, \quad \mbox{otherwise} \end{array}\right. , 
\qquad\quad
Y^i_{(b)} = \left\{\begin{array}{c}\, y^i, \quad i \in\{3,4\} \\ 0, \quad \mbox{otherwise} \end{array}\right. ,  
\eea
while all other components vanish.

Let us recall the oscillator representation of the corresponding $\R^2_{\theta_a}$. It is given by the ladder operators
\bea
y^1 + i y^2 &=& a^\dagger, \qquad y^1 - i y^2 = a, \qquad [a,a^\dagger] = 2 \Theta^{12} =: \theta_a,\\
y^3 + i y^4 &=& b^\dagger, \qquad y^3 - i y^4 = b, \qquad [b,b^\dagger] = 2 \Theta^{34} =: \theta_b ,
\label{y-oscillator}
\eea
which act on the Hilbert spaces $\cH_{a}$ and $\cH_{b}$ respectively and they satisfy\footnote{Note that the sign of 
$\theta$ should be considered as positive in these formulae.}
\bea
y^1 y^1 + y^2 y^2 &=& \frac 12(a^\dagger a + a a^\dagger) = \theta_a(\hat n_a + \frac 12),   \nn\\
y^3 y^3 + y^4 y^4 &=& \frac 12(b^\dagger b + b b^\dagger) = \theta_b(\hat n_b + \frac 12)   .
\eea
Here $\hat n_a = a^\dagger a$ and $\hat n_b=b^\dagger b$ are the usual occupation number operators.

We also introduce a fermionic oscillator representation for the Gamma matrices\footnote{We always assume a basis
where the $\R^{2n}_\theta$ has canonical form as in \eq{y-oscillator}.},
\bea
2\a &=& \Delta_1 - i \Delta_2, \quad 2\a^\dagger =  \Delta_1 + i \Delta_2, 
\qquad  \{\a,\a^\dagger\} = 1,  \nn\\
2\b &=& \Delta_3 - i \Delta_4, \quad 2\b^\dagger =  \Delta_3 + i \Delta_4, 
\qquad  \{\b,\b^\dagger\} = 1,  \nn \eea
with $\{\a,\b\} = 0$ and we define the following chirality operators,
\bea 
\chi_{\a} &=& i\Delta_1\Delta_2 = -2(\a^\dagger \a-\frac 12),  \qquad \chi_{\b} =  i\Delta_3\Delta_4 =  -2(\b^\dagger\b-\frac 12),\nn 
\eea
acting on the spin-$\frac 12$ irreducible representation. Moreover, it is straightforward to show that
\bea\Sigma_{12} &=& \frac i4 [\Delta_1,\Delta_2] = \frac 12[\a,\a^\dagger] = \frac 12(1-2\a^\dagger\a) =  \frac 12\chi_{\a} 
\eea and similarly $\S_{34}=\frac 12 \chi_b$.
Note that the chirality operator on $\R^2_{a}\times \R^2_{b}$ is then given by $\chi = \chi_{\a} \chi_{\b}$.

The most general state $\Psi_{(ab)}$ can  be written as 
\be
\Psi_{(ab)} = |00\rangle \psi_{00} + |01\rangle \psi_{01}
 + |10\rangle \psi_{10} + |11\rangle \psi_{11},
\label{general-fermion-state}
\ee
where the ket denotes spinor states in obvious notation.
Now recall \eq{block-algebra}, which implies that 
the most general wave-function $\psi \in \cH_a \otimes \cH_b^*$ can be written in terms of the eigen-basis 
of the correpsonding harmonic oscillator algebras,
\be
\psi \in \oplus_{n,m\geq 0} \, |n\rangle_a\langle m|_b,
\ee
for any given spinor component.
Thus the Dirac operator  for the off-diagonal spinors  can be written as 
\bea
\slashed{D}_6 \Psi _{(ab)} &=& \Delta_i (Y^i_{(a)} \Psi_{(ab)} - \Psi_{(ab)} Y^i_{(b)}) \nn\\
&=& (\a a^\dagger + \a^\dagger a)\Psi_{(ab)} - (\b\Psi_{(ab)} b^\dagger + \b^\dagger \Psi_{(ab)} b) \nn\\
&\stackrel{!}{=}& 0,
\label{Dirac-action-offdiag-general}
\eea where in the last line we demand the Dirac operator to vanish so as to capture the corresponding zero modes. Moreover,
\bea
\slashed{D}_6^2 \Psi _{(ab)} &=& \Box \Psi_{(ab)} + \Sigma [\Theta,\Psi_{(ab)}] \nn\\
 &=& \theta_a(\hat n_a+\frac 12)\Psi_{(ab)} 
+ \Psi_{(ab)}\theta_b(\hat n_b + \frac 12)  
 +  \Sigma  \Theta_{(ab)}\Psi_{(ab)} \nn\\
&=& \theta_a \Big(\hat n_a + \frac 12 (1-\chi_{\a})\Big) \Psi_{(ab)}
 + \theta_b \Big(\hat n_b +\frac 12 (1+\chi_{\b})\Big) \Psi_{(ab)}. 
\label{D2-stretch-EV}
\eea
Observe that
$Y^i_{(a)} \Psi_{(ab)} Y^i_{(b)} = 0$ for the perpendicular branes under consideration.
Here we define the effective flux 
\be
\Theta_{(ab)} := [\Theta,.]|_{\Psi_{(ab)}} = \Theta_{(a)} - \Theta_{(b)},
\label{off-diag-theta}
\ee
acting on the off-diagonal modes $\Psi_{(ab)}$.

It follows that there is a single zero mode, which satisfies $\hat n_a = \hat n_b = 0$ 
and $\chi_{\a} = 1, \, \chi_{\b} = -1$. 
This implies that 
\be
\Psi_{(ab)}^{0,0} = |10\rangle \psi_{10}^{0,0}, \qquad 
  \psi_{10}^{0,0} =  |0\rangle_a\langle 0|_b,
\label{localized-state}
\ee
which is naturally interpreted as bound state localized at the intersection of the two branes.
This state is manifestly chiral, since $\chi_{\a} = - \chi_{\b} = 1$ and hence $\chi = -1$.
It corresponds to a fermionic ``string`` connecting the two branes.
In addition, there is a tower of massive fermions 
\be
\Psi_{(ab)}^{n,m} = |ij\rangle \psi_{ij}^{n,m}, \qquad 
  \psi_{ij}^{n,m} =  |n\rangle_a\langle m|_b.
\label{localized-state-tower}
\ee
These can be interpreted as long fermionic strings stretching between the branes.

The chiral zero modes can be identified more quickly if we assume localization right away, so that 
$\psi_{ij} = |0\rangle_a\langle 0|_b$ in \eq{general-fermion-state}. Then 
\bea
\slashed{D} \Psi _{(12)} &=& \a a^\dagger\Psi_{(12)} - \b^\dagger \Psi_{(12)} b  = 0, 
\eea
which gives again \eq{localized-state}.
Clearly this mechanism generalizes to any (orthogonal) intersection
of quantum planes.

Let us note that in the present case of $\R^2\cap\R^2 \subset \R^6$,
there are two remaining transversal directions. Therefore
this mechanism will not lead to chiral zero modes in
the ten-dimensional model. In order to avoid this problem, we will consider intersections of higher-dimensional branes which span the entire $\R^6$. 
In that case, we will indeed obtain the desired chiral fermions.

\paragraph{Generalization.}

Since the $Y^i_{(a)}$ will always generate quantum planes $\R^{2n}_{(a)}$ in the extra dimensions, we can 
generalize the above oscillator construction by choosing
a basis such that $\R^{2n}_{(a)}$ decomposes canonically as products of 2-dimensional quantum planes
 $\R^{2n}_{(a)} = \R^{2}_{\theta_1} \times ... \times \R^{2}_{\theta_n}$.
Rather than writing down cluttered general formulae, we discuss a few more cases explicitly.

\subsubsection{$\R^2 \cap \R^4$}
\label{sec:D5D7}

In order to span the full $\R^6$, let us now consider a system of a $D5$ brane $D_a$ and a $D7$ brane
$D_b$, embedded along $\R^2_{45} \cap \R^4_{6789}$. 
Generalizing the setup of the previous section, 
we need to add another bosonic algebra with operators $b',b'^{\dagger}$, satisfying $[b',{b'}^\dagger] = \theta_{b'}$, for $\R^4_{6789}$, as well as a set of fermionic operators $\b',\b'^\dagger$ with
$\{\b',{\b'}^\dagger\} = 1$, for the Clifford algebra. The Dirac operator becomes
\bea
 \slashed{D}_6 \Psi _{(ab)} &=& \Delta_i (Y^i_{(a)} \Psi_{(ab)} - \Psi_{(ab)} Y^i_{(b)}) \nn\\
 &=& (\a a^\dagger + \a^\dagger a)\Psi_{(ab)} - (\b\Psi_{(ab)} b^\dagger + \b^\dagger \Psi_{(ab)} b + \b'\Psi_{(ab)} {b'}^\dagger + {\b'}^\dagger\Psi_{(ab)} b' ) 
\nn
\eea
and moreover,
\bea
\slashed{D}_6^2 \Psi _{(ab)}
 &=& \theta_a(\hat n_a+\frac 12)\Psi_{(ab)} 
+ \Psi_{(ab)}\theta_b(\hat n_b + \hat n_{b'} + \frac 12)  
 +  \Sigma  \Theta_{(ab)}\Psi_{(ab)} \nn\\
&=& \theta_a \Big(\hat n_a + \frac 12 (1-\chi_{\a})\Big) \Psi_{(ab)}
 +  \Big(\theta_b(\hat n_b +\frac 12 (1+\chi_{\b}))
   +  \theta_{b'} (\hat n_{b'} +\frac 12 (1 +\chi_{\b'}))  \Big) \Psi_{(ab)} . \nn
\eea
This implies again that all zero modes are localized as
$\psi = |0\rangle_a\langle 0,0|_{b}$, and have the form
\be
\Psi_{(ab)}  = |100\rangle |0\rangle_a\langle0,0|_{b} .
\ee
This is clearly a chiral mode in $\R^6$ with $\chi = +1$, which implies also chirality in the 4-dimensional space-time 
because of the Majorana-Weyl condition in ten dimensions. 
If the two branes are replaced by $n_a$ resp. $n_b$ 
coinciding branes, we obtain chirally protected massless fermions on $\R^4$ transforming as
$(n_a) \otimes (\obar n_b)$ under $U(n_a) \times U(n_b)$. These are the
building blocks for the low-energy physics in the matrix model framework.

Note that
in the case of two branes $D_{(a)}\cap D_{(b)}$ intersecting perpendicularly along a common $\R^4_\theta$, the chirality
of the localized zero mode is given by 
\be
\chi(\Psi _{(ab)}) = (-1)^{d_{b}/2} ,
\ee
where $d_{b}$ denotes the number of extra dimensions of the brane $D_{(b)}$. In particular,
for the above example we have
$\chi(\Psi _{(ab)}) =1$, while $\chi(\Psi _{(ba)}) =-1$.

\subsubsection{$K \cap \R^2$}

In principle, one can consider intersections of compact non-commutative branes such as 
fuzzy spheres  and tori $K = S^2_N$, $K = T^2_N$ \cite{Madore:1991bw}. 
These can be easily realized in terms of block-matrix configurations as before. 
From a semi-classical point of view one would expect that 
chiral zero-modes  arise similarly from localized off-diagonal fermions.
However, this is more complicated because of curvature. Indeed, the compact space may intersect the other 
brane in more than one location.
We will appeal to the expected qualitative features in some examples below but 
 postpone a more careful investigation of such scenarios to future work.

\subsubsection{$\R^4 \cap \R^4$}
\label{sec:R4R4}

It is also interesting to consider the system of two $D7$ branes, e.g. $\R^4_{4567} \cap \R^4_{6789}$. This system contains an additional common $\R^2$. Therefore, if the 
non-commutative structure respects this sub-space, the above story goes through with 
the obvious modifications. It leads to a single chiral zero mode 
with chirality $\chi = +1$ on the (4+2)-dimensional intersection $\R^6$. 
However, from a physical point of view, we do not want flat $\R^6$ but rather chiral zero-modes on 4D space-time.

There are two obvious ways out. First, we can assume that the 6-dimensional intersection
has non-trivial geometry such as $\R^4 \times T^2$ or $\R^4 \times S^2$.
We can then put a flux on this compact space\footnote{this is achieved if the two copies of 
the compact space $K$ on the two branes have different quantization numbers $N$, e.g. $S^2_N$ and $S^2_{N-1}$ respectively
\cite{Steinacker:2003sd,Chatzistavrakidis:2009ix,Aoki:2010hx,Karabali:2001te}; 
see also e.g. \cite{Aschieri:2006uw} for related work.}, 
which implies (via the index theorem) that 
the lowest Kaluza-Klein modes on $K$ are chiral and massless.
This leads to a chiral effective 4D action, and the number of zero-modes is determined by the
flux on $K$.
Such compactified extra dimensions are very reasonable physically, and 
we will apply this mechanism in section \ref{standard-model}.

There is another possibility which should also be explored. Even though  
$\R^4_{4567} \cap \R^4_{6789} = \R^2_{67}$, it may be that the non-commutative structures $\theta^{ab}$ 
do not respect this 
sub-space. Then the above derivation must be modified, which should be studied elsewhere.

\subsubsection{$\R^2 \cap \R^2 \cap \R^4$}

As a further example consider $\R^2_{45} \cap \R^2_{67} \cap \R^4_{6789}$.
This is a system of one $D7$ and two $D5$ branes. According to the above results,
$\Psi_{(12)}$ connecting the $\R^2_{45}$ and the $\R^2_{67}$ brane is not chiral. 
Similarly $\Psi_{(23)}$ connecting the $\R^2_{67}$ and $\R^4_{6789}$ is also non-chiral, 
since in that case there are two common directions and the full six-dimensional space is not saturated. 
However, the component $\Psi_{(13)}$ connecting $\R^2_{45}$ and $\R^4_{6789}$ is  chiral.

In the low-energy effective action on the common $\R^4$, we expect that non-chiral fermions will acquire
a dynamical mass and disappear from the low-energy spectrum. Only the chiral zero modes are protected. 
Those are candidates for the fermions of the standard model.


In order to verify the above expectations let us consider the Dirac operator on the spinors. Evidently,
\bea 
\slashed{D}_6\Psi_{(12)}&=&(\a a^{\dagger}+\a^{\dagger}a)\Psi_{(12)}-(\b\Psi_{(12)}b^{\dagger}+\b^{\dagger}\Psi_{(12)}b), \\
\slashed{D}_6\Psi_{(13)}&=&(\a a^{\dagger}+\a^{\dagger}a)\Psi_{(13)}-(\b\Psi_{(13)}b^{\dagger}+\b^{\dagger}\Psi_{(13)}b
+\b'\Psi_{(13)}{b'}^{\dagger}+{\b'}^{\dagger}\Psi_{(13)}b'), \\
\slashed{D}_4\Psi_{(23)}&=& -(\b'\Psi_{(23)}{b'}^{\dagger}+{\b'}^{\dagger}\Psi_{(23)} b'),
\eea
noting that $\R^2_{67} \cap \R^4_{6789}$ corresponds to a 6-dimensional intersection,
leaving only 4 extra dimensions.
Accordingly, the square of the Dirac operator on the modes may be determined and subsequently be set to zero. As a result we find
\bea 
\slashed{D}_6^2\Psi_{(12)} &=& 0 \quad \Rightarrow \,{\chi_{\a}=1, \chi_{\b}=-1, \chi_{\b'}=\pm1}, \\
\slashed{D}_6^2\Psi_{(13)} &=& 0\quad  \Rightarrow \,{\chi_{\a}=1, \chi_{\b}=-1, \chi_{\b'}=-1}, \\
\slashed{D}_4^2\Psi_{(23)} &=& 0 \quad \Rightarrow \,{\chi_{\a}=\pm1, \chi_{\b}=\pm1, \chi_{\b'}=-1} ,
\eea 
where $\chi = \pm 1$ indicates that the corresponding chirality is undetermined.
According to this result, only $\Psi_{(13)}$ is chiral in 4D, which  confirms the above expectations.

\subsection{Remarks on supersymmetry}
\label{sec:susy}

Let us briefly discuss the supersymmetry of the brane configurations under consideration here. 
The basic brane solutions \eq{basic-brane-2n} of the matrix model are known to be $\frac 12$ BPS states,
preserving the supersymmetry \eq{susy}  $\d^{(1)}_\e + \d^{(2)}_\xi$  for 
$\xi = \frac 12 \Theta^{ab}\Gamma_{ab}\e$.
Thus 2 intersecting branes preserve a supersymmetry if and only if 
$\xi = \frac 12\Theta^{ab}_{(1)}\Gamma_{ab}\e = \frac 12\Theta^{ab}_{(2)}\Gamma_{ab}\e$, hence
\be
\Theta^{ab}_{(12)}\Gamma_{ab}\,\e = 0
\ee
using the notation \eq{off-diag-theta}. 
This means that $\Theta^{ab}_{(LR)}\Gamma_{ab}$ has reduced rank. 
Not surprisingly, this quantity will also play a significant role in the effective interaction 
between the branes as discussed in section \ref{sec:one-loop}. 
Denoting with $f_i$ the eigenvalues of $\Theta^{ab}_{(12)}$ as in Appendix A, the eigenvalues of
$\Theta^{ab}_{(12)}\Gamma_{ab}$ in the eigenbasis \eq{dirac-basis} 
are given by $\a_{n_i} = \sum_{n_i=\pm 1} n_i f_i$. 
Since we assume in this paper that $\Theta^{ab}$ for the two branes coincides on the intersection, 
it follows that $\Theta^{ab}_{(12)}$ has at least 2 vanishing eigenvalues, 
and the remaining $f_i, i=3,4,5$ are determined by the noncommutative flux in the extra dimensions.

Now consider the case of 2 intersecting $D5$ branes as in section \ref{sec:D5D5}. Then $f_3$ and $f_4$ are the 
extra fluxes on the 2 branes, while $f_5=0$. Thus half of the $\a_{n_i}$ are zero if and only if $f_3 = \pm f_4$.
In that case these configurations are $\frac 14$ BPS.
Indeed, we will see in section \ref{sec:one-loop} that this is precisely the case where the interaction vanishes. 
An analogous result holds in the case of 2 intersecting $D7$ branes.
This may be expected in view of known results 
in the literature \cite{Gauntlett:1997cv,Kim:2000mp}. 
In the case of  $D5\cap D7$ intersecting in a $D3$ as in section \ref{sec:D5D7},
the eigenvalues $\a_{n_i} = \sum_{n_i=\pm 1} n_i f_i$ for given $f_3,f_4,f_5 \neq 0$ 
can vanish only for special choices of fluxes, such that e.g. $f_3+f_4=\pm f_5$. In such cases, the
configuration is $\frac 18$ BPS, but generically there is no supersymmetry.
Indeed we will see in section \ref{sec:one-loop} that the interaction may have either sign in this case.

\section{Towards realistic scenarios}

Intersecting $D$-brane models proved to be a very fruitful arena in the quest of embedding the SM in string theory. 
Indeed, there exists a vast literature on the ongoing exploration of type II orientifold vacua which 
aspire to successfully describe the SM or Grand Unified Theories (GUTs) thereof 
(for reviews and a more complete list of references see \cite{Blumenhagen:2006ci,Marchesano:2007de}).

In the present section we would like to discuss the possibility of realistic model building in the context of matrix models, 
based on the results of the previous sections. 
Let us first explain what is meant here by realistic. The minimum requirements that we shall impose include:
\begin{itemize} 
\item The SM gauge group at low energies (plus some additional $U(1)$ factors which become massive).
\item Chiral fermion spectrum.
\item Correct hypercharge assignment.
\end{itemize}
Another obvious requirement is that the model is anomaly free, which is discussed in section
\ref{sec:anomalies}.
Having ascertained these requirements, one may subsequently try to impose
more phenomenological or theoretical requirements, such as proton stability, 
gauge coupling unification, 
family replication, mass hierarchies etc. Such a systematic analysis is left for future work. 
In the present work our main goal is to show that realistic model building in the above sense is indeed possible 
within the matrix model framework. 

The most economic way to obtain the SM from intersecting branes in the matrix model appears to be 
via  four branes  $D_a, D_b, D_{c}, D_{d}$, which carry the gauge groups $U(3)_C$, $U(2)_L$, $U(1)_c$ and $U(1)_d$.
This possibility has been explored extensively in the context of string theory,
with several possible variations, cf. \cite{Ibanez:2001nd,Blumenhagen:2005mu,Blumenhagen:2006ci,Marchesano:2007de}.
We will discuss such configurations in the matrix model below.
However, there are important differences to the string theory approach.
The main difference is that we  consider branes with extra dimensions embedded in $\R^{10}$, rather than
compactifing the 10-dimensional target space as in string theory\footnote{The point is that 
embedded NC branes form a natural class of solutions in matrix models, 
and the IKKT model is expected to be UV finite on such backgrounds 
(with compactified extra dimensions such as fuzzy spheres), 
as in $\cN=4$ SYM theory.}. 
This has several implications. First, there is no tadpole condition, due to the non-compact embedding space.
On the other hand, there are more restrictions for the representation content than in the string theory approach. 
In particular, all fermions must be in the $(n_i)\otimes (\obar n_j)$
of a pair of $U(n_i)$ gauge groups, and no  $(n_i)\otimes (n_j)$ can arise.
We will be led to essentially a single configuration with the correct spectrum of chiral zero-modes.
This is essentially equivalent to a 
configuration of non-intersecting branes which was found previously
in the matrix model framework \cite{Grosse:2010zq}, however without a mechanism for chirality.
This mechanism is provided now by the intersections. 

We will also discuss a possible generalization of this brane configuration with five branes, 
which allow to obtain right-handed neutrinos. 


\subsection{Standard model from four intersecting branes}
\label{standard-model}

Consider four branes  $D_a, D_b, D_{c}, D_{d}$ with gauge group 
\bea
\label{gg1a} G&=&U(3)_C\times U(2)_L\times U(1)_c\times U(1)_d, \nn\\ 
&=&SU(3)_C\times SU(2)_L\times U(1)_a\times U(1)_b\times U(1)_c\times U(1)_d, 
\eea 
It seems plausible that all intersections should be ''close to each other``, so that the branes
can be considered as flat to a good approximation; distant intersections would presumably 
lead to decoupled sectors. 
We use this as a working hypothesis here, which allows us to work with planar branes.
Furthermore, we restrict ourselves to the case of perpendicular intersections as above.

Clearly the left-handed quarks $Q_L$ must arise on $D_a \cap D_b$, the right-handed quarks $u_R, d_R$ on $D_a \cap D_c$ 
and $D_a \cap D_d$ respectively, say, and the right-handed electron $e_R$  on 
$D_c \cap D_d$.
Thus all of these intersections must exist and provide localized chiral fermions. 
It is clear that this cannot work if $D_a$ is a $D5$ brane\footnote{if $D_a$ is a $D5$ brane embedded e.g. along
$\R^2_{45}$, then all the $D_b, D_c$ and $D_d$ would have to be embedded along $\R^4_{6789}$, and 
there would be no chiral leptons.}.
Furthermore, a doublet of left-handed leptons $l_L$ should arise from the intersection of $D_b$ with one of the two 
$U(1)$ branes (this turns out to be possible, which is a non-trivial statement); let us denote this
one with $D_d$. 
We will see that {\em no} second doublet of left-handed ''exotic`` leptons will arise at the intersection of $D_b$
with the other $U(1)$ brane.

We can therefore assume that $D_a$ is a $D7$ brane embedded along  $\R^4_{4567}$. 
In order to accommodate $Q_L, u_R$ and $d_R$, all other branes must contain $\R^2_{89}$. 
Furthermore $D_b$ and $D_d$ must span together the entire $\R^6$ in order to 
provide  $l_L$ from $D_b \cap D_d$, therefore they must both be 7-branes.
Thus up to equivalence, we can assume that $D_b$ is  embedded along  $\R^4_{6789}$ and
$D_d$ is embedded along $\R^4_{4589}$. Finally, in order to obtain $e_R$ from $D_c \cap D_d$
it follows that $D_c$ is embedded along  $\R^4_{6789}$. Thus we arrive at the embedding in
table 1, with the particles at the intersections given in table 2.
\begin{table}\begin{center}
    \begin{tabular}{|c|c|c|}\hline \multicolumn{3}{|c|}{Table 1}\\
    \hline
   brane &  gauge group & brane embedding  \\ \hline
   $D_a$ &  $U(3)_C$ & D7 along  $\R^4_{4567}$     \\ \hline
   $D_b$ &  $U(2)_L$ & D7 along  $\R^4_{6789}$     \\ \hline
   $D_c$ &  $U(1)_c$ & D7 along  $\R^4_{6789}$     \\ \hline
   $D_d$ &  $U(1)_d$ & D7 along  $\R^4_{4589}$     \\ \hline
   \end{tabular} 
\caption{Gauge group and brane embedding for the model with four intersecting branes.}
\end{center}\end{table}
\begin{table}\begin{center}
    \begin{tabular}{|c|c|c|c|}\hline \multicolumn{4}{|c|}{Table 2}\\
    \hline
    Intersection & Representation & Particle & flux \\ \hline
    $D_a\cap D_b$ & $(\bar 3,2)(-1,1,0,0)$ & $Q_L$ & $N_\b'-N_\b$
     \\ \hline
    $D_a\cap D_c$ & $(\bar 3,1)(-1,0,1,0)$ & $d_R$ & $N_\b''-N_\b$ \\ \hline
    $D_a\cap D_d$ & $(\bar 3,1)(-1,0,0,1)$ & $u_R$ & $N_\a'-N_\a$ \\ \hline
 	$D_d\cap D_b$ & $(1,2)(0,1,0,-1)$ & $ l_L$ & $N_\g-N_\g''$  \\ \hline
	$D_d\cap D_c$ & $(1,1)(0,0,1,-1)$ & $e_R$  & $N_\g'-N_\g''$ \\ \hline 
   \end{tabular}
\caption{Particle spectrum at the brane intersections. The representation 
content appearing in the second column is given for the gauge group 
$SU(3)_C\times SU(2)_L\times U(1)_a\times U(1)_b\times U(1)_c\times U(1)_d$, 
where the first parenthesis contains the quantum numbers for the non-abelian factors and the second parenthesis 
the abelian charges in the above order. The fourth column is related to the fluxes seen by the fermions 
on the compact spaces $K$, as discussed in the text.}
\end{center} \end{table}
This corresponds to the following 
matrix realization of the chiral matter as anticipated in \cite{Grosse:2010zq},
\be
\Psi=\begin{pmatrix}
  0_{2} & { \begin{array}{cc} 0 &  l_L \end{array}} & Q_L \\
   &  \begin{array}{cc}  0 &  e_R \\  & 0 \end{array} & Q_R \\
   &    & 0_{3}
\end{pmatrix},
\label{particle-assign}
\ee
with
\bea
Q_L = \begin{pmatrix} u_L \\ d_L   \end{pmatrix}, \qquad
l_L = \begin{pmatrix} \nu_L \\ e_L   \end{pmatrix}, \qquad
 Q_R = \begin{pmatrix}
   d_R \\ u_R
\end{pmatrix}
 \label{matrixPsi-2}.
\eea
Here the branes are arranged in the order $D_b, D_c, D_d,D_a$, which is of course just conventional. 
The correct hypercharge is then reproduced by  
\bea
Y &=& \begin{pmatrix}
   0_{2}  &  &  &  \\
   & - \sigma_3 & &  \\
   & &  & -\frac 13 \one_{3}
\end{pmatrix}  
 = - \frac 13 Q_a- Q_c+ Q_d, \label{Ygenerator}  
\eea
which acts in the adjoint; here $Q_{a,b,c,d}$ denotes the $U(1)$ charges of the branes.
Note that it follows from the above discussion that the intersection $D_b \cap D_c$ is not chiral, therefore
there are {\em no} exotic chiral leptons with representation\footnote{notice that these are the quantum numbers of 
the second Higgs doublet of the MSSM.} $(1,2)(0,1,-1,0)$. 
The lower-diagonal blocks of the fermionic matrices are related to the upper-diagonal blocks 
by the 10-dimensional Majorana condition, and not displayed here.
They correspond to the anti-particles with conjugate representation content.

Let us discuss the chirality in more detail. Since  all branes are $D7$ branes, 
all zero modes on the 6-dimensional pairwise intersections have 6-dimensional 
chirality $\chi=+1$, as discussed in  section \ref{sec:R4R4}.
This holds both for the upper-diagonal and the lower-diagonal entries in  \eq{particle-assign}. 
In order to get 4-dimensional chiral fermions, we assume that the 
intersections $D_i \cap D_j$ are compactified e.g. on 2-dimensional  tori or spheres with a flux $m_{ij}$. 
Via the index theorem,
this leads to $|m_{ij}|$ fermionic zero modes $\psi_{(ij)}$ with chirality ${\rm sign}(m_{ij})$.
Note that the opposite zero modes $\psi_{(ji)}$ see the opposite flux $m_{ji} = - m_{ij}$, and 
therefore have opposite chirality. Hence we need to realize the fluxes $m_{ij}$ such that the correct
4-dimensional chiral fermion content of the standard model is obtained. 

We now give such a flux compactification which reproduces the standard model. 
Take $D_a$ to be $K_{\a} \times K_{\b}$, $D_b$ to be $K_{\b}' \times K_{\g}$, 
$D_c$ to be $K_{\b}'' \times K_{\g}'$,
and  $D_d$ to be $K_{\a}' \times K_{\g}''$ (always omitting the common $\R^4$). 
Here $K$ stands for either $T^2_N$ or $S^2_N$, and 
the subscripts indicate the tangent plane at the location of the intersection.
Then the geometries of the intersections are as follows
 $D_a\cap D_b \cong K_{\b} \cong D_a\cap D_c$,  $D_a\cap D_d \cong K_{\a}$, 
 $D_b\cap D_d \cong K_{\g} \cong D_c\cap D_d$. 
However, e.g. $K_{\a}$ and $K_{\a}'$ can have different quantization parameters 
$N_\a$ and $N_\a'$, which implies that 
e.g. $\psi_{(ab)}$ feels a flux $N_\b -N_\b'$ while $\psi_{(ad)}$ feels a flux $N_\a -N_\a'$. 
Thus a possible assignment which gives the correct number of generations and chiralities is
\begin{align}
N_\g-N_\g'' &= 3, \quad N_\g'-N_\g'' = -3,  \nn\\
N_\b''-N_\b &= -3, \quad N_\b'-N_\b = 3,  \nn\\
N_\a'-N_\a &= - 3 .
\end{align}
It is obvious that this does have solutions. In this way, we have found a background which reproduces the 
exact standard model spectrum in the matrix model at low energies, including the appropriate number of generations.

In general, compactified extra dimensions may intersect more than once. This would lead to additional 
hidden sectors, or possibly to additional generations. Since the generations are realized above through indices
associated to fluxes, we do not want any additional intersections here. We can indeed provide a
realization of the above $K_\a, K_\b, K_\g$ as 2-spheres with the desired properties. 
This is achieved e.g. by the following embedding 
\begin{align}
 & K_\a \cong S^2 \subset \R^3_{456} \quad  \mbox{centered at} \quad   \vec e_6,  \nn\\
 & K_\b \cong S^2 \subset \R^3_{678} \quad   \mbox{centered at} \quad   \vec e_8,  \nn\\
 & K_\g \cong S^2 \subset \R^3_{894}  \quad  \mbox{centered at} \quad   \vec e_4 . 
\label{sphere-embedding}
\end{align}
Here $\vec e_i$ denotes the unit vector $(0,...,1,...,0)$ in direction $i$.
 It is then easy to check that these spheres have only a single pairwise intersection
at the origin, with tangent space $\R^2_{45}$, $\R^2_{67}$, $\R^2_{89}$, respectively.

Such fuzzy sphere solutions indeed exist in the matrix model \cite{Iso:2001mg,Kimura:2001uk}, e.g.
upon adding appropriate cubic terms to the action. The  spheres \eq{sphere-embedding}
can be realized by adding a term of the form 
$\Tr \big(\varepsilon_{abd}^{(456)} X^a X^b X^c \, + \, \varepsilon_{abd}^{(678)} X^a X^b X^c \, + \, 
 \varepsilon_{abd}^{(894)} X^a X^b X^c \big)$, 
where the superscript in  e.g. $\varepsilon_{abd}^{(456)}$ indicates the possible values of the indices.
It is easy to check that the desired fuzzy spheres are then indeed solutions of the matrix model.

Of course, adding such explicitly symmetry breaking terms to the model is undesirable
from the gravity point of view.
As shown in \cite{Steinacker:2011wb}, there are in fact analogous solutions of the IKKT model 
{\em without} any additional terms, by giving them angular momentum and a 
modified NC structure.
Although their intersections have not been studied, one should expect that the same qualitative 
features arise; this will be elaborated elsewhere.

Finally, we recall that  all branes were assumed to be approximately flat
and intersecting at $\pi/2$.
This assumption may be too strong, and there may be different realizations of the SM upon relaxing these conditions.
In particular if the branes intersect at different loci on the compactified extra dimensions, 
then it may be possible to consider also e.g. combinations of $D5$ and $D7$ branes. 
Intersections at different locations may lead to  hidden sectors or to family replication.
We leave such non-trivial geometries for future work.
Nevertheless, it is quite striking that the correct SM spectrum arises quite naturally for
the above configuration of intersecting branes, without exotic chiral particles.

\subsection{Other models from five intersecting branes}
\label{fivebranemodels}

Previously we showed that in order to realize the SM within the MM framework, 
the minimal set-up consists of four intersecting branes, which all have to be $D7$ ones. Thus the MM necessarily singles out one configuration, where the branes have six-dimensional intersections. This fact led us to the compactification of the two extra dimensions and the introduction of fluxes on them in order to achieve the correct SM spectrum. 

The importance of the above restriction to $D7$ branes may be further illuminated by asking 
whether {\it{realistic}} brane configurations with purely four-dimensional intersections can be constructed in the present framework. The answer is no 
and the reason is the following. The obvious way to avoid having 
six-dimensional intersections is to relax the consideration of {\it{four}} intersecting branes. 
Indeed, let us consider a single fifth brane, say $D_e$, carrying a gauge group $U(1)_e$. 
Now the color brane can be either $D5$ or $D7$, since the additional freedom introduced by the presence of the fifth 
brane allows to circumvent the argument of the previous section. Then there 
are essentially two ways (up to equivalence) of embedding the branes in $\R^{10}$, 
involving only 4-dimensional intersections of $D5$ with $D7$ branes as presented in table 3. 
Chiral particles may be accommodated at the intersections as in table 4. 
 On the other hand, the intersections $D_a\cap D_e,D_b\cap D_c,D_b\cap D_d$ and $D_c\cap D_d$ are not chiral in this model 
and therefore {\it {chiral}} particles associated with them do not exist. 
This leads to the correct representation content of the SM (with right-handed neutrino), however the chiralities 
come out wrong. The point is that here we have no freedom to add fluxes in extra dimensions,
which could fix the chiralities as in the previous section.


\begin{table}
\begin{center}
    \begin{tabular}{|c|c|c|c|}\hline \multicolumn{4}{|c|}{Table 3}\\
    \hline
   brane &  gauge group & brane embedding 1 & brane embedding 2  \\ \hline
   $D_a$ &  $U(3)_C$ & D7 along  $\R^4_{4567}$ &  D5 along  $\R^4_{45}$    \\ \hline
   $D_b$ &  $U(2)_L$ & D5 along  $\R^2_{89}$ & D7 along  $\R^4_{6789}$    \\ \hline
   $D_c$ &  $U(1)_c$ & D5 along  $\R^2_{89}$  &  D7 along  $\R^4_{6789}$ \\ \hline
   $D_d$ &  $U(1)_d$ & D5 along  $\R^2_{89}$   & D7 along  $\R^4_{6789}$ \\ \hline
   $D_e$ &  $U(1)_e$ & D7 along  $\R^4_{4567}$ & D5 along  $\R^4_{45}$ \\ \hline	
   \end{tabular}
\caption{Gauge group and brane embeddings for the models with five intersecting branes.}
\end{center}
\begin{center}
    \begin{tabular}{|c|c|}\hline \multicolumn{2}{|c|}{Table 4}\\
    \hline
    Intersection & Representation   \\ \hline
    $D_a\cap D_b$ & $(\bar 3,2)(-1,1,0,0,0)$ 
     \\ \hline
    $D_a\cap D_c$ & $(\bar 3,1)(-1,0,1,0,0)$  \\ \hline
    $D_a\cap D_d$ & $(\bar 3,1)(-1,0,0,1,0)$   \\ \hline
 	$D_b\cap D_e$ & $(1,2)(0,1,0,0,-1)$  \\ \hline
	$D_c\cap D_e$ & $(1,1)(0,0,1,0,-1)$  \\ \hline
        $D_d\cap D_e$ & $(1,1)(0,0,0,1,-1)$  \\ \hline
   \end{tabular}
\caption{Particle spectrum at the brane intersections. The representation 
content appearing in the second column is given for the gauge group 
$SU(3)_C\times SU(2)_L\times U(1)_a\times U(1)_b\times U(1)_c\times U(1)_d\times U(1)_e$.}
\end{center}\end{table}

According to the above, our framework is very restrictive and 
forces us to consider $D7$ branes. In the present case of five branes this can 
easily be realized by appropriately promoting the 
$D5$ branes of table 3 to $D7$ ones and adding appropriate fluxes on the intersections, as before. 
This is interesting because the particle corresponding to the last row of table 4 
is naturally identified with a right-handed neutrino $\nu_R$. Therefore the extension 
of the SM by a $\nu_R$ is achieved in our framework with five $D7$ branes.
Then the upper triangular matrix realization is
\be
\Psi=\begin{pmatrix}
  0_{2} & 0 & { \begin{array}{cc} 0 &  l_L \end{array}} & Q_L \\
   &  0  & \begin{pmatrix}
            0 &  e_R \\  0 &  \nu_R 
           \end{pmatrix}
 & Q_R \\
 & & ~~~~0 & 0 \\
&  &  & ~0_{3}
\end{pmatrix},
\label{particle-assign2}
\ee
while the correct hypercharge assignment is given by 
\bea
Y &=& \begin{pmatrix}
   0_{2\times 2}  &  &  & & \\
   & - \sigma_3 & & & \\
   & & & 1 & &	\\
   & & & & -\frac 13 \one_{3\times 3}
\end{pmatrix}  
 = - \frac 13Q_a- Q_c+ Q_d+ Q_e. \label{Ygenerator2}  
\eea

\subsection{Bosonic sector}

So far, we only considered fermions in the off-diagonal blocks.
There are clearly also 
bosonic matrices connecting the branes, corrersponding to scalar fields i.e. (generalized) Higgs fields.
However, most of them will be massive and disappear from the low-energy physics. To see this, consider their action 
in the above $\R^2 \cap \R^2$ background obtained from \eq{Box-off-diag}, which takes the form
\be
\Box \phi_{(ab)} = \theta_a(\hat n_a+\frac 12) \phi_{(ab)} + \phi_{(ab)} \theta_b(\hat n_b+\frac 12),
\label{Box-bosons-stretching}
\ee
because $Y^i_{(a)} \phi_{(ab)} Y^i_{(b)} = 0$ vanishes in that background.
The minimum is clearly obtained for $\hat n_a \phi_{(ab)} = 0 = \phi_{(ab)} \hat n_b$, thus
\be
\phi_{(ab)}^{0,0} =  |0\rangle_a\langle 0|_b .
\ee
However, these lowest modes now have a non-vanishing  mass
\be
M_0^2 \phi_{(ab)}^{0,0} = \frac 12 (\theta_a+\theta_b) \phi_{(ab)}^{0,0} .
\label{mass-scalar}
\ee
The fact that this mass does not vanish is due to the uncertainty in their localization, which 
leads to a quantum mechanical ''zero-point energy`` determined by the 
NC scale $\theta$. In particular, supersymmetry is broken by non-commutativity.


The Yukawa couplings for the chiral fermions have the form $\obar\Psi_{(ba)}  \phi_{(bc)} \Psi_{(ca)}$,
which involves three branes. 
Since these bosonic modes $\phi_{(bc)}$ are massive for intersecting branes, the Higgs field responsible for 
electroweak symmetry breaking can arise only from off-diagonal blocks connecting parallel branes.
Indeed there are two parallel branes $D_b$ and $D_c$ in the first model in section \ref{standard-model},
and the bosonic mode connecting these branes has the correct quantum numbers of the 
electroweak Higgs.
It is easy to check that this leads to valid Yukawa couplings for the SM,
however the couplings of the $u_R$ quarks vanishes (since the bosonic mode connecting the 
$D_b$ and the $D_d$ brane is massive). 
In the second model  in section \ref{fivebranemodels}, there are 3 parallel branes 
$D_b,D_c$ and $D_d$, which may accomodate a rich Higgs sector and 
allows to realize all the necessary Yukawa couplings.
Notice also that there are always several Higgs fields within a given block,
corresponding to the different matrix components $Y^i$ in the internal space. 
Their interactions should originate from the $[Y^i,Y^j]^2$ term in the matrix model, although the precise mechanism
is at present unclear.


\subsection{Anomaly and other issues}
\label{sec:anomalies}

In string theory, there is a further constraint on the intersecting branes due to  
the tadpole cancellation condition \cite{Blumenhagen:2006ci}. 
It results from consistency conditions on the fluxes in compactified extra dimensions.
However for branes embedded in flat $\R^{10}$ as considered here, 
this does not lead to any further constraints. 

In the IIB matrix model, the only possible constraint on the brane configurations 
comes from stability considerations, notably 
at one loop. We study the stability of the intersecting brane configurations below. Although we cannot 
give a full analysis for the case of compactified branes, we do find clear evidence that intersecting
branes can have an attractive interaction and thus form bound states. 

Another important issue is the (chiral) anomaly. 
In the intersecting branes scenario, there are $U(1)$ factors arising on each brane.
The overall trace-$U(1)$ is part of the gravity sector. 
Certain combinations of them combine to form anomalous-free $U(1)$'s, in particular as required in the
standard model. However, some $U(1)$'s typically are anomalous from the low-energy 
point of view. On the other hand, it is clear that 
the entire model (including all massive brane modes arising in the off-diagonal sectors)
is free of anomalies. This means that the low-energy anomalies are not pathological, but
lead to non-standard implementation of the corresponding symmetry. 

One mechanism which arises in a similar field-theoretical context is the St\"uckelberg mechanism \cite{Kors:2005uz},
which implies that these would-be massless anomalous 
gauge fields acquire a mass and thereby disappear from the low-energy
spectrum. This is related to the generalized Green-Schwarz mechanism \cite{Blumenhagen:2006ci} which governs the 
analogous issue in the string theory context. 
The precise implementation of these mechanisms in the present model should be studied in more detail 
elsewhere. Here we simply refer to the consistency of the model at the fundamental level and 
assume that these would-be anomalous $U(1)$'s disappear from the low-energy sector of the model.

Another important issue is the stability of these brane configurations, in particular
the identification of brane configurations which are bound states and do not collapse into coinciding branes. 
The discussion of the 
one-loop effective action in the next section shows that there are several mechanisms which 
govern the interaction of such branes, depending on their fluxes, relative orientation, and so on.
This is clearly a complicated dynamical issue which will require much more time 
and work to be understood. 
We will set up a suitable formalism for such an analysis, 
and briefly discuss some qualitative aspects.

Finally, it should be clear that what we obtained here is a version of the standard model which lives
on noncommutative rather than commutative space-time. This means that there is a scale 
determined by $\theta^{\mu\nu}$
where the fuzzyness of the branes under consideration becomes important, both 
for space-time as well as for the internal spaces. Due to the maximal supersymmetry of the 
IIB model (corresponding to $\cN=4$ SUSY on a $\R^4_\theta$ brane), the pathological 
UV/IR mixing effects are expected to be absent \cite{Matusis:2000jf}, 
but noncommutativity plays a central role in the gravity sector \cite{Steinacker:2010rh}.
It is therefore reasonable to expect that the physics of the model reduces to that of an
ordinary gauge theory at low energies. However, more work is required to fully understand the
impact of noncommutativity here.

\section{One-loop effective action and stability}
\label{sec:one-loop}

The most important aspect of the matrix model is that there is a clear concept of quantization:
one should simply integrate over the space of all (bosonic and fermionic) matrices. 
This is the matrix analog of the Feynman path integral.
In fact, the one-loop effective action for a general given background $X^a$
in the matrix model can be written down in a remarkably compact 
way\footnote{Additional Wess-Zumino-type terms  may arise in fully non-degenerate 
10-dimensional backgrounds. They will not be discussed here, although they may be
relevant in the most interesting configurations with chiral zero-modes, leading to non-trivial
magnetic interactions \cite{Tseytlin:1999tp}.} \cite{Ishibashi:1996xs},
\begin{align}
\Gamma_{{\rm 1-loop}}[X] 
&= \frac 12 \Tr \(\log(\Box  + \Sigma^{(Y)}_{ab}[\Theta^{ab},.])
-\frac 12 \log(\Box  + \Sigma^{(\psi)}_{ab}[\Theta^{ab},.])
- 2 \log \Box\)   \nn\\
&= \frac 12  \( \Tr_{(10)}\log(\one  + \Sigma^{(10)}_{ab}\Box^{-1}[\Theta^{ab},.])
-\frac 12 \Big( \Tr_{(16)}\log(\one  + \Sigma^{(16)}_{ab}\Box^{-1}[\Theta^{ab},.])\) \nn\\
 &= \frac 12 \Tr \Bigg( -\frac 14 (\Sigma^{(10)}_{ab} \Box^{-1}[\Theta^{ab},.] )^4 
 +\frac 18 (\Sigma^{(16)}_{ab} \Box^{-1}[\Theta^{ab},.])^4 \,\, +  \cO(\Box^{-1}[\Theta^{ab},.])^5  \Bigg). \nn\\
\label{Gamma-IKKT}
\end{align}
Here the traces are taken over operators on $Mat(\infty,\C)\otimes V$, where $V$ is either the 10-dimensional vector
or the 16-dimensional spinor representation of $SO(9,1)$, and
\begin{align}
\Theta^{ab} &=-i[X^a,X^b] \, , \nn\\
(\Sigma_{ab}^{(16)})^\a_\b &= \frac i4 [\Gamma_a,\Gamma_b]^\a_\b,\nn\\
(\Sigma_{ab}^{(10)})^c_d &= i(\d^c_a g_{bd} - \d^c_b g_{ad})\, , \nn\\
\Box \phi &= [X^a,[X_a,\phi]]\, .
\end{align}
It is easy to see using $SO(10)$ group theory that the first three terms in the Taylor expansion
in $\Theta^{ab}$ cancel identically, reflecting the maximal SUSY. The leading non-trivial term 
in such an expansion is given by the last line in \eq{Gamma-IKKT}.
This implies that there are no UV divergences 
for fluctuations around 4-dimensional {\em and} 6-dimensional flat branes at one loop\footnote{ 
For 8-dimensional branes, there are log-divergences for non-compact branes, but one may
expect that there are no divergences even for 8-dimensional NC branes 
provided they are compactified.}.

The one-loop effective action is expected to capture the leading behaviour of branes in IIB 
supergravity \cite{Ishibashi:1996xs}. 
We will briefly indicate how to apply this formula in 
the case of intersecting brane configurations. This should allow to understand their stability.

For the backgrounds under consideration here (i.e. blocks of various $\R^{2n}_\theta$), 
we observe that $[\Box,[\Theta,.]] = 0$. Therefore the 1-loop effective action can be written neatly 
in exponentiated form using a Schwinger parametrization\footnote{This is based on 
the identity $\int_0^\infty \frac {ds}{s} (e^{-s A} - e^{-s B}) = \ln B - \ln A$. },
\begin{align}
&\Gamma_{\!\textrm{1loop}}[X]\! = - \frac 12 \Tr \int\limits_0^\infty \frac {ds}{s} 
   \Big( e^{-s(\Box  + \Sigma^{(Y)}_{ab}[\Th^{ab},.])}
  - \frac 12 e^{-s(\Box + \Sigma^{(\psi)}_{ab}[\Th^{ab},.]) }  - 2 e^{-s \Box}  \Big) \nn\\
&= - \frac 12 \Tr \int\limits_0^\infty \frac {d s}{s}  e^{-s\Box}
   \Big( e^{-s \Sigma^{(10)}_{ab}[\Th^{ab},.]}
  - \frac 12 e^{-s\Sigma^{(16)}_{ab}[\Th^{ab},.] }  - 2  \Big) \nn\\
&= - \frac 12  \int\limits_0^\infty \frac {d s}{s} \Tr_\cA (e^{-s\Box})
   \Big( \tr_{(10)} (e^{-s\Sigma^{(10)}_{ab}[\Th^{ab},.]})
  - \frac 12 \tr_{(16)} ( e^{-s\Sigma^{(16)}_{ab}[\Th^{ab},.] })  - 2  \Big) .
\label{finite-N-IKKT}
\end{align}
Here we separated the wave functions into a $SO(10)$ sector with trace $\tr$ over the 
 appropriate  (vector or spinor)
representation $V$, and a space-time sector denoted by $\cA \cong Mat(\infty,\C)$. 
The latter consists of the modes around the background $X^a$.

For the $SO(10)$ part, note that any given constant $\Theta^{ab}$ is in one-to-one 
correspondence\footnote{To see this, 
we can bring it into standard form involving only $2\times 2$ antisymmetric block matrices on the diagonal, 
i.e. such that it is an element of the Cartan algebra. 
Using the Killing metric, this defines a corresponding weight $\mu$.} with 
a generator of $SO(10)$. Hence the $SO(10)$ structure enters only via the characters 
\begin{align}
\chi_V(H) &=  Tr_{V}(e^{H}), 
\end{align}
which are invariant under the adjoint action as well as Weyl reflections. 
This takes care of the $SO(10)$ sector for fixed $[\Theta_{ab},.]$.

For a single flat brane $\R^{2n}_\theta$, the one-loop effective action vanishes because $[\Theta^{ab},.] = 0$,
and the characters cancel due to SUSY.
Now consider a background of two flat branes as in \eq{2-branes},
denoted as $\R^{2n}_L$ and $\R^{2n'}_R$ to be specific.
We organize the space of modes accordingly into block structure.
The contribution due to the diagonal blocks still vanishes as before.
However we now get non-trivial contributions from the off-diagonal blocks, where 
$\Theta_{(LR)} = [\Theta,.] = \Theta_{(L)} - \Theta_{(R)}$ 
is the difference of the ``NC flux`` between the branes. 
This contribution is given by the following simple formula
\begin{align}
&\Gamma_{\!\textrm{1loop}}[X]\! = - \frac 12  \int\limits_0^\infty \frac {d s}{s} \Tr_\cA (e^{-s\Box})
   \chi(-s\Th_{(LR)}),
\label{loop-IKKT-chi}
\end{align}
where
\begin{align}
\chi(s\Theta_{(LR)}) &:= \chi_{10}(s\Theta_{(LR)}) - \frac 12  \chi_{16}(s\Theta_{(LR)}) - 2 \nn\\
 &= \tr \Big(e^{s\Sigma^{(10)}_{ab}\Th^{ab}_{(LR)}  }\Big)
  - \frac 12 \tr \Big( e^{s\Sigma^{(16)}_{ab}\Th^{ab}_{(LR)} }\Big)  -2  .
\end{align}
Therefore the characteristic function $\chi(\Theta_{(LR)})$ governs the effective interaction between the branes. 
In particular, the sign of the one-loop action is determined by
the sign of $\chi(\Theta_{(LR)})$.

These characters can be evaluated explicitly. We already pointed out that the first three terms in the Taylor expansion
for $\chi(s\Theta)$ vanish identically; this is a consequence of maximal SUSY.
Moreover, it is not hard to show \cite{1-loop-MM} 
that the leading contribution in $\Theta$ to the 1-loop action 
has the form
\begin{align}
\chi(s\Theta)|_{O(\Theta^4)} 
 \sim s^4 \(\tr ((\Theta g)^4) - \frac 14 (\tr(\Theta g)^2)^2\) \,\, + O(s\Theta^5), 
\label{character-asymptotics}
\end{align}
where $\Theta g$ is viewed as $10\times 10$ matrix.
If $\Theta$ has rank 4, then this term is positive definite \cite{Tseytlin:1999dj},
\be
\tr ((\Theta g)^4 - \frac 14 (\tr(\Theta g)^2)^2 \sim  (\Theta-\star_g \Theta)^2 (\Theta+\star_g \Theta)^2 
\equiv \Theta_+^2 \Theta_-^2 \,\, \geq 0 ,
\label{rank4-positive}
\ee
where $\star_g$ is the Hodge dual with respect to the relevant 4-dimensional metric $g_{ab}$.
Here $\Theta_\mp$ denotes the (anti-)self-dual components of $\Theta$ in its 4-dimensional subspace, 
which vanishes if and only if $\Theta$ is (anti-)self-dual. We will in fact derive a stronger result below.
Thus the effective potential $W = \Gamma_{\rm 1-loop} \sim -\chi <0$ is generically attractive
in the rank 4 case. This is consistent with 
the results of \cite{Ishibashi:1996xs} for the interaction of two anti-parallel D1-branes, which 
have rank two.

\paragraph{Explicit results for the characters.}

We summarize some results for the characters as explained in the appendix A.
For a given flux $\Theta_{LR}^{ab}$, 
we can choose a basis using a suitable 
$SO(9,1)$ rotation where $\Theta_{LR}^{ab}$ is block-diagonal:
\begin{align}
\Theta_{LR}^{ab} = \begin{pmatrix} 0 & f_1 & & & \\
                       -f_1 & 0  & & &  \\
                   & & \ddots & & \\
                      &  & & 0 & f_5 \\
                      &  & & -f_5 & 0
 \end{pmatrix}
= \begin{pmatrix} f_1 \, i\sigma_2 & & \\
                   & \ddots  & \\
                        &  & f_5\,i\sigma_2 \\
 \end{pmatrix}
. 
\end{align}
Notice that, in general, this basis need not coincide with the basis adapted to the intersection of the branes.
For the vector representation this gives
\begin{align}
\chi_{(10)}(s\Theta_{LR}) =\tr(e^{s\Theta_{LR}^{ab}\Sigma^{(Y)}_{ab}}) &= \sum_{i=1}^5 (e^{2s f_i} +  e^{-2s f_i}) 
\,, \label{Vector-character}
\end{align}
while for the spinor representation one finds 
\begin{align}
2\chi_{(16,\pm)}(s\Theta_{LR}) 
 = \chi_{(32)}(s\Theta_{LR}) &= (e^{s f_1} + e^{-s f_1}) \ldots (e^{s f_5} + e^{-s f_5})
\,, \label{Dirac-character}
\end{align}
provided the rank is no more than 8
(since then both contributions from $e^{\pm\a f_5}$ coincide).

Furthermore, we show in appendix A that 
\begin{align}
\chi(s\Theta_{LR}) &= (e^{\frac{s}2 (f_1-f_2)} - e^{- \frac{s}2 (f_1-f_2)})^2(e^{ \frac{s}2 (f_1+f_2)} - e^{-\frac{s}2 (f_1+f_2)})^2 \geq 0,
 & \mbox{rank} \,\Theta_{LR} \leq  4  \nn\\
\chi(s\Theta_{LR}) & \quad ... \, \mbox{either sign}, & \mbox{rank} \,\Theta_{LR} \geq 6 . \nn
\end{align}
The first statement implies \eq{rank4-positive},
since $f_1 = \pm f_2$ is equivalent to $\Theta_{LR}$ being (anti)self-dual.
Note that if the rank of $\Theta_{LR}$
is 6 or higher, then $\chi(\Theta_{LR}) > 0$ if two eigenvalues dominate but do not coincide, but
 $\chi(\Theta_{LR}) < 0$ if all eigenvalues are comparable.

\subsection{Interaction between branes}

Since the bare matrix model action does not give any interaction between two branes, the one-loop effective action
should give the correct interaction at leading order.

Consider two branes along $\R^{2n}_a$ and $\R^{2n'}_b$. As shown above, the sign of their
effective potential  $W = \Gamma_{\rm 1-loop} \sim -\chi(\Theta_{LR})$ is governed by $\chi(\Theta_{LR})$. 
If $\Theta_{LR}$ has rank $\leq 4$ then there is an attractive interaction, which cancels in the (anti-)self-dual case.
Therefore low-dimensional branes should tend to coincide. 
If ${\rm rank} \Theta_{LR}>4$, then branes may repel each other, in particular if
the eigenvalues of $\Theta_{(LR)}$ approximately coincide. However  vacuum states 
should have an overall attractive interaction, which arises if two eigenvalues dominate but do not coincide.
The attractive interaction is maximized if there are two dominating eigenvalues of $\Theta_{(LR)}$.

For example, consider a $D3$ and a $D5$ brane which intersect in $\R^4$, i.e. the $D3 \subset D5$.
If their fluxes coincide in the common $\R^4$, then $\Theta_{(LR)}$ has rank 2 and we have a 
bound state. If the $D3$ and $D5$ have a generic orientation relative to each other, 
the rank of $\Theta_{(LR)}$ would typically be 6, and the interaction tends to be repulsive. Therefore
the former brane configuration is preferred.

Now consider two $D5$ branes. Again for a generic relative orientation, 
the rank of $\Theta_{(LR)}$ would  be large and the interaction tends to be repulsive.
However if they intersect in $\R^4$ with identical fluxes, then $\Theta_{(LR)}$ has rank 4,
and they form again a bound state. This shows that intersecting brane configurations are not academic 
artefacts but preferred bound states. Moreover, the NC structures along the space-time $\R^4$
indeed prefer to (almost) coincide, while the attraction is mediated by the flux in the extra dimensions.


In the  case of a $D5$ brane and a $D7$ brane, the situation is more complicated. However, 
intersecting configurations should again be preferred, such that the flux cancels along their intersection.
If the remaining $\Theta_{(LR)}$ is dominated by two eigenvalues, a bound state should form again. 
However since $\Theta_{(LR)}$ must have rank 6 in order to have chiral fermions, a more complete analysis 
including e.g. also the Wess-Zumino-type terms would be required.
A similar discussion applies to the case of two intersecting $D7$ branes.
This shows that the configurations of interest here may indeed be natural vacua of the matrix model.

In  configurations with multiple branes,
there might be a competition between the attractive and repulsive interactions
between various brane pairs. One may hope that  the physically relevant brane configurations
are favoured in this way, leading possibly even to the required negative mass for the physical Higgs
(since there are different forces which act on the branes). 
In fact, it is clear from \eq{loop-IKKT-chi} that the 1-loop interactions discussed here are indeed due to the 
various off-diagonal modes which connect the branes.


Another class of bound states arises if the flux on the common $\R^4$ dominates but is 
not identical for the two branes, so that $\Theta_{(LR)}$ is dominated by the rank 4 sector along
$\R^4$. This arises for e.g. anti-parallel branes. 
Such a scenario is interesting because Lorentz-violating effects on space-time $\R^4$ due to $\Theta^{\mu\nu}$
may be averaged out.

\paragraph{Standard model compactification.}

Now consider the realization of the standard model in section \ref{standard-model}
in more detail, with the specific 
compactification and fluxes as discussed there.
Consider the intersection of $D_a$ with $D_b$, say.
The intersection is on a  2-sphere $K_\b$ with net flux $\Theta_{\rm intersect} \sim N_\b' - N_\b = 3$.
This is much smaller than the individual fluxes $\sim N_\a, N_\g$ on the two other, (locally) perpendicular spheres 
$K_\a$ resp. $K_\g$ of the branes. Therefore the character $\chi(\Theta_{LR})$ which governs the 
interaction is dominated by the (difference of the) fluxes on $K_\a$ and $K_\g$, while the 
flux from $K_\b - K_\b'$ can be neglected. This means that we are indeed in the situation discussed above where
the intersecting branes form a bound state. The same observation applies to all other pairwise intersections,
except between $D_b$ and $D_c$. Therefore this type of vacuum indeed forms a tightly bound state!

Of course there will be other configurations which also form bound states, and we cannot yet argue that this
standard model vacuum is the preferred one. But it certainly is a  reasonable candidate, 
and it is very remarkable 
that we {\em can} indeed ask and possibly answer such questions in a meaningful way.

\subsection{Kinetic sector}

Finally, we discuss how to evaluate the trace over the kinetic term $\Tr_\cA(e^{-\a\Box})$.
We will only consider the case where both branes have a common $\R^4_\theta$ sector, which commutes with the 
remaining extra-dimensional $\R^{2n}$. Then
\be
\Box = \Box_4 + \Box_6, \qquad [\Box_4,\Box_6] = 0,
\ee
so that $e^{-\a\Box} = e^{-\a\Box_4} e^{-\a\Box_6}$. 
Then the ''scalar`` wave-functions for $\Psi_{\pm}$ live in 
\be
\cH_{\pm} = \cA_{\cM^4}\otimes |n_a,n_b\rangle\langle n_c|,
\ee
noting that any spin dependence is captured by $\chi(\a\Theta)$.
The trace over the space-time modes is easy to carry out,
\be
\tr_{\cA}(e^{-\a\Box_4}) = \int\frac{d^4 p}{(2\pi\LNC^2)^2} e^{-\a\LNC^{-2} p\cdot p} 
= \frac 1{\a^2} ,
\ee
where $\LNC^{-4} = \sqrt{|\theta^{\mu\nu}|}$ is the scale of noncommutativity on $\R^4_\theta$.
The off-diagonal oscillator modes give the following contribution 
\begin{align}
\tr (e^{-\a\Box_6}) &=  \sum_{n_a,n_b,n_c=0}^\infty e^{-\a (\theta_a n_a + \theta_b n_b + \theta_c n_c)}  
  = \frac 1{(1-e^{-\a\theta_a})(1-e^{-\a\theta_b})(1-e^{-\a\theta_c})} \nn\\
 &= \left\{\begin{array}{ll} 1+ O(e^{-\a\theta}), & \a\to\infty \\
                             \a^{-3}, & \a \to 0
           \end{array}\right.
\label{box-6-sum}
\end{align}
if $\Theta^{ab}_{LR}$ has  rank 10, and similarly for lower rank. 
Here  $\theta = \min(\theta_i)$. 
Note that the $n_i=0$ mode corresponds to the localized zero (or almost-zero) modes, while the 
$n_i>0$ correspond to the excited massive ''stretched string`` modes. 
The pole at $\a=0$ only arises because of the infinite sum over the $n_i$. 
This leads to a UV divergence of the 1-loop effective action for rank $>6$. 
However in the (more realistic) case of compactified extra dimensions, e.g. on a fuzzy sphere $K=S^2_N$, there 
are typically only finitely many Kaluza-Klein-like modes labeled by $n_i$,  
and no such UV divergence appears. We therefore assume that the extra dimensions are compactified, 
and replace the rhs of \eq{box-6-sum} by $1+ O(e^{-\a\theta})$. 
%
Then
\be
\Gamma = \int_0^\infty \frac{d\a}{\a^3}\, \chi(\a\Theta) (1+ h(\a\theta)) .
\ee
Recall also that $\chi(\a\Theta) = O(\a^4)$, therefore the integrals have no UV divergence. 
However, the zero modes  lead to a (standard) IR divergence $\a\to\infty$. 
This arises because the  fields in the off-diagonal terms are charged and therefore 
contribute in the loop, unlike the block-diagonal modes.
This may be taken care of by  subsequent spontaneous symmetry breaking, notably in the standard model.

In any case, 
it should be kept in mind that the backgrounds under consideration here are somewhat special.
More general cases with fluxes on the branes that mix $\R^4$ with the extra dimensions will be 
treated elsewhere, but the qualitative features are expected to survive.

\section{Discussion and conclusions}

The main objective of the present paper was to explore the possibility of 
constructing realistic models for particle physics within the type IIB matrix model. 

We first identified solutions of the matrix model  
corresponding to multiple flat non-commutative branes, which moreover intersect 
with each other. The fermion spectrum of the model in such 
backgrounds was studied in detail. It was shown that chiral fermionic 
zero-modes arise at the intersection of the branes, provided these branes span 
together the full ten-dimensional space $\R^{10}$.

Having established the existence of chiral modes in the above backgrounds, 
we initiated a search for realistic scenarios in this context.
In particular, we presented a brane configuration which provides
 a realization of the standard model at low energies. It consists of four mutually intersecting 
$D7$ branes, which accommodate the particles of the SM at their 
intersections. It is worth noting that in the present context, the fact that 
all the branes are $D7$ is not an arbitrary choice but it is imposed by the 
requirement of correct embedding of the SM. Indeed, had one or more of the four branes been $D5$ instead, some of the chiral particles of the SM would 
have not appeared in the model. This is a first manifestation of the fact 
that the present framework is more restrictive than similar constructions in 
string theory. 
Moreover, it comes as a bonus that the model does not contain any 
exotic chiral particles. Furthermore, we should note that in the present model the intersections between 
the $D7$ branes are six-dimensional, and therefore some compactification is required. 
Compactified solutions, such as fuzzy spheres and fuzzy tori, are indeed known to exist in the matrix model. 
We have shown that one can 
indeed construct a workable model assuming that the extra dimensions are 
compactified. In addition, it turned out that the presence of appropriate fluxes in 
these extra dimensions is necessary in order to obtain the desired SM 
spectrum. It is welcome that these fluxes can also account for the 
number of generations in the SM. 

Clearly, different brane configurations lead to different models. For 
example, models including a right-handed neutrino may be constructed in the 
matrix model upon the addition of a fifth abelian brane, as it was shown in 
section \ref{fivebranemodels}. It is worth noting that the additional freedom 
due to the introduction of a fifth brane allows in principle for the construction of brane configurations with purely four-dimensional intersections 
i.e. without six-dimensional ones. However, these fail to reproduce the 
correct SM spectrum without additional compactifications and fluxes. Therefore, 
even in this case $D7$ brane configurations are necessary. This is 
another manifestation of the fact that the present framework is very restrictive. This fact could be considered as an advantage, since it means that the matrix 
model offers less freedom on what can be done and what cannot and therefore 
favours less models than in similar string constructions.

Another interesting point is that supersymmetry is broken for the configuration considered here, 
due to the fluxes on the intersecting noncommutative branes. The reason is that while chiral fermions are 
protected, the bosonic modes aquire a mass due to the uncertainty relations on the NC branes. 

Let us stress that an important feature of the above considerations and models is that they also 
include gravity. 
Indeed, if the IKKT model describes the type IIB string theory in the non-perturbative regime, 
then its solutions should also capture features associated to gravity. 
Moreover, since F-theory corresponds 
to type IIB string theory at strong coupling \cite{Vafa:1996xn}, the IKKT model should also capture 
mechanisms associated to F-theory, such as the recently conjectured
relations with non-commutative geometry  \cite{Heckman:2010pv,Cecotti:2009zf,Furuuchi:2010gu}. 
%

Finally, the issue of stability of the above configurations was addressed using the one-loop effective action.
We argued that for certain flux configurations, intersecting branes may form a bound state 
and therefore do not collapse into coinciding branes. This is argued to hold for the standard model 
realization presented in this paper. Therefore the matrix model framework allows to address the difficult
questions about the vacuum structure and its low-energy effective action in a meaningful way. 
We have demonstrated through an explicit construction  
that the standard model may indeed arise within this framework.
The phenomenological aspects of this realization should be studied in more detail in future work.

\paragraph{Acknowledgments.}

We thank F. Lizzi for collaboration in the early stages of this project.
H.S. would like to thank D. Blaschke for a related collaboration 
on the one-loop effective action, 
which provided valuable insights. Useful discussions with R. Blumenhagen, O. Ganor,  S. Iso, H. Kawai,  Y. Kitazawa,  D. L\"ust,  
J. Nishimura and N. Sasakura are greatfully acknowledged, as well as hospitality at UC Berkeley, KEK and the Yukawa Institute. 
A.C. would like to thank the Department of Physics of the University of Vienna for hospitality.
This work was partially supported by the SFB-Transregio TR33
"The Dark Universe" (Deutsche Forschungsgemeinschaft), the European Union 7th network program 
"Unification in the LHC era" (PITN-GA-2009-237920), and the NTUA's programmes supporting basic research PEBE  2009 and 2010.
The work of H.S. was supported by the Austrian Science Fund (FWF) under contract
P21610-N16.

\bigskip

\bigskip
\newpage

{\LARGE{\textbf{Appendix}}}

\appendix


\section{Characters}
\label{app:characters}

The character for a representation $V$ of some Lie algebra $\mg$ is defined as 
\be
\chi_V(H) = \tr\, e^{H} 
\,, 
\ee
where $H\in \mg$ (which is often assumed to be in the Cartan subalgebra and thus identified with a weight).
Characters are very useful objects in group theory, notably because they satisfy
$\chi_{V \otimes W} = \chi_V \chi_W$. In the present context, 
we can interpret the term $\tr e^{\a \Sigma_{ab}\Theta^{ab}_{LR}}$
as character of $SO(10)$.

For a given flux $\Theta_{LR}^{ab}$ (e.g. in the semi-classical limit at some point),  we can choose a basis 
using a suitable $SO(10)$ rotation where $\Theta_{LR}^{ab}$ is block-diagonal:
\begin{align}
\Theta_{LR}^{ab} \sim \begin{pmatrix} 0 & f_1 & & & \\
                       -f_1 & 0  & & &  \\
                   & & \ddots & & \\
                      &  & & 0 & f_5 \\
                      &  & & -f_5 & 0
 \end{pmatrix}
= \begin{pmatrix} f_1 \, i\sigma_2 & & \\
                   & \ddots  & \\
                        &  & f_5\,i\sigma_2 \\
 \end{pmatrix}.
\label{theta-standard} 
\end{align}
In this basis, we can then choose a corresponding  fermionic oscillator rep. for the Gamma matrices,
\begin{align}
2\a_i &= \Gamma_{2i-1} - i \Gamma_{2i}\,, \qquad 2\a_i^\dagger =  \Gamma_{2i+1} + i \Gamma_{2i}\,, \qquad
\{\a_i,\a_j^+\} = \d_{ij}\,, \quad i=1,2,...,5 \,, \nn\\
i\Gamma_1\Gamma_2 &= -2(\a_1^+ \a_1-\frac 12)\,,  \qquad
 i\Gamma_3\Gamma_4 =  -2(\a_2^+\a_2-\frac 12)\,, \,\mbox{etc.} \nn\\
\Sigma_{12} &= \frac i4 [\Gamma_1,\Gamma_2] = \frac 12[\a_1,\a_1^+] = \frac 12(1-2\a_1^+\a_1) =  \frac 12\chi_{(1)}
= \frac 12 \sigma_3
\, \quad \mbox{etc.}, 
\end{align}
which act on the spin $\frac 12$ irrep.
Thus
\begin{align}
\Theta_{LR}^{ab}\Sigma^{ab} &= \sum_i f_i \,\one \otimes ... \otimes \sigma_3 \otimes ... \otimes \one , \label{sigma-rep}\\
\chi_{(32)}(\a\Theta_{LR}) = \tr_{32}(e^{\a\Theta_{LR}^{ab}\Sigma^{(\psi)}_{ab}}) &= (e^{\a f_1} + e^{-\a f_1}) \ldots (e^{\a f_5} + e^{-\a f_5})
 = \sum_{n_i=\pm 1} e^{\a n_i f_i} 
\,, 
\end{align}
which acts on $\C^{32}$, where the $\sigma_3$ in \eq{sigma-rep} is in the $i$-th tensor slot.
The most general state for a Dirac fermion $\Psi$ can  be written as 
\be
\Psi = \sum_{n_i = \pm 1}  \psi_{n_1 \ldots n_5} |n_1 \ldots n_5\rangle \psi_{n_1 \ldots n_5} ,
\label{dirac-basis}
\ee
where the ket denotes spinor states.
It is the sum of both chiral representations, whose character is
\begin{align}
&\chi_{(16,+)}(\a\Theta_{LR}) = \sum_{n_i=\pm 1,\sum n_i = 5,1,-3}  e^{\a n_i f_i} \nn\\
 &= e^{\a (f_1+f_2+f_3+f_4+f_5)}+  e^{\a (f_1+f_2-f_3-f_4+f_5)} + e^{\a (f_1+f_2-f_3-f_4-f_5)} + e^{\a (f_1+f_2+f_3-f_4-f_5)}  \nn\\
&\quad +  e^{\a (-f_1-f_2+f_3+f_4+f_5)} + e^{\a (-f_1+f_2-f_3+f_4+f_5)} + e^{\a (-f_1+f_2+f_3-f_4+f_5)} + e^{\a (f_1+f_2+f_3+f_4-f_5)} \nn\\
&\quad +  e^{\a (f_1-f_2-f_3+f_4+f_5)} + e^{\a (f_1-f_2+f_3-f_4+f_5)} + e^{\a (f_1-f_2+f_3+f_4-f_5)} \nn\\
&\quad + e^{\a (f_1-f_2-f_3-f_4-f_5)} + e^{\a (-f_1+f_2-f_3-f_4-f_5)} + e^{\a (-f_1-f_2+f_3-f_4-f_5)}\nn\\
&\quad + e^{\a (-f_1-f_2-f_3+f_4-f_5)} + e^{\a (-f_1-f_2-f_3-f_4+f_5)} \,, 
\,\label{Weyl-character}
\end{align}
and similarly for $\chi_{(16,-)}$.
Notice that $\tr_{16,\pm}(e^{\a\Theta_{LR}^{ab}\Sigma^{(\psi)}_{ab}}) = \frac 12 \tr_{32}(e^{\a\Theta_{LR}^{ab}\Sigma^{(\psi)}_{ab}})$
whenever $\Theta_{LR}$ has rank at most 8, since then both contributions from $e^{\pm\a f_5}$ coincide.

On the vector representation $(10)$, we can use the $10 \times 10$ matrix 
$(\Theta^{ab}_{(LR)})\d_{bb'}$  itself as generator\footnote{for general 
signature one should replace $(\Theta_{(LR)}\d)$ with $(\Theta_{(LR)}\eta)$.
However only the Euclidean directions will be
relevant below, thus we will ignore this distinction.},
i.e.
\be
\cJ_{(LR)} := \Sigma^{(\rm vector)}_{ab}\Th^{ab}_{(LR)} \equiv (\Theta_{(LR)})^{ab} .
\ee
This gives
\begin{align}
\chi_{(10)}(\a\Theta_{LR}) =\tr(e^{\a\Theta_{LR}^{ab}\Sigma^{(Y)}_{ab}}) &= \sum_i (e^{2\a f_i} +  e^{-2\a f_i}) 
\,. 
\end{align}
Thus for a general rank 4 flux one finds
\begin{align}
\chi(\a\Theta_{LR}) &= \tr_{10}(\a\Theta_{LR}) - \frac 14 \tr_{32}(\a\Theta_{LR}^{ab}) - 2
= (e^{2\a f_1} + e^{-2\a f_1} +e^{2\a f_2} + e^{-2\a f_2} +6) \nn\\
  &  - \frac 14 8 (e^{\a f_1} + e^{-\a f_1})(e^{\a f_2} + e^{-\a f_2})  - 2 \nn\\
&= (e^{\a (f_1-f_2)/2} - e^{-\a (f_1-f_2)/2})^2(e^{\a (f_1+f_2)/2} - e^{-\a (f_1+f_2)/2})^2\nn\\
& \geq 0\ ,
\end{align}
which is positive definite, and vanishes precisely for (anti-) self-dual 4-dimensional fluxes $\Theta_{LR}$. 
This is consistent with previous results.

For a 6- and higher-dimensional flux, $\chi$ can have either sign. 
Let 
\be
A_i = (e^{\a f_i} + e^{-\a f_i}) = 2 \cosh(\a f_i)\geq 2 .
\ee
Clearly $\chi(\Theta_{LR})$ is a totally symmetric polynomial in the $A_i$. More precisely,
\begin{align}
\chi(\Theta_{LR}) &= (A_1^2  + A_2^2 + A_3^2 -2) -A_1 A_2 A_3 - 2, & \mbox{rank} \,\,\Theta_{LR} = 6 , \nn\\
\chi(\Theta_{LR}) &= (A_1^2  + A_2^2 + A_3^2 + A_4^2-6) - \frac 12 A_1 A_2 A_3 A_4 - 2, 
& \mbox{rank} \,\,\Theta_{LR} = 8 . \nn
\end{align}
Clearly if there are two 
dominating eigenvalues $f_{1,2}$ and the remaining $|f_i|$ are much smaller, 
then $\chi \approx \chi(f_1,f_2) \geq 0$. However, they become negative if three of more eigenvalues 
become comparable\footnote{note that for given $A_i \geq 2$, the sum $A_i^2+A_j^2$ is minimized for fixed $A_i A_j$ if and only if $A_i = A_j$.
Therefore $\chi(\Theta_{LR})$ takes its minimum for given product of the $A_i$ if and only if all $A_i$ coincide.}. 
Indeed for coinciding eigenvalues $A_i \approx A$ in the rank 6 case, 
\be
\chi^{(3)}\approx
\chi^{(3)}(A) = 3 A^2-A^3-4 = -(-2 + A)^2 (1 + A) \leq 0 .  
\ee
Similarly for the rank 8 case,
\be
\chi^{(4)}\approx
\chi^{(4)}(A) = 4 A^2-\frac 12 A^4 - 8 = - \frac 12 (-4 + A^2)^2  \leq 0, 
\ee
which become 
negative since the fermionic sector dominates. 
The rank 10 case is more complicated since we need to distinguish between the chiralities, but clearly
again either sign can arise.

\section{Supersymmetry and the zero-modes}

\paragraph{Supersymmetric quantum mechanics}



The oscillator approach, which was used in section 3 to describe the quantum planes, naturally leads to algebraic structures which are well-known from Quantum Mechanics (QM). In particular, it will be shown here that there is a natural realization of a supersymmetry algebra, in analogy to supersymmetric QM (for an excellent review see \cite{Cooper:1994eh}; see also chapter 3 of \cite{Denef:2011ee}). In fact supersymmetry provides an alternative framework to study the existence of fermionic zero-modes due to its well-known relation to the index theorem \cite{Witten:1982df,AlvarezGaume:1983at,Friedan:1983xr}.

A supersymmetric quantum mechanical system consists of a Hilbert space ${\cal H}$, which separates into two subspaces ${\cal H}_b$ and ${\cal H}_f$, 
 associated with bosonic, say $|b\rangle$, and fermionic, say $|f\rangle$,  states respectively. 
On this Hilbert space there is an action of the operators $H$, the Hamiltonian, and $Q^i,Q^{i\dag},i=1,\dots N$, 
the supercharges, satisfying the following supersymmetry algebra{\footnote{This is the $sl(1/1)$ superalgebra.}},
\bea \{Q^i,Q^{j\dag}\} &=& 2\d^{ij}H, \nn \\
        {[}Q^i,H{]} = [Q^{i\dag},H] &=& 0, \nn \\
        (Q^i)^2 = (Q^{i\dag})^2 &=& 0. \label{susyalg}\eea
Moreover, the fermion number operator or Witten index $(-1)^F$, characterizing the spin-statistical nature of each quantum state, 
may be defined. 
This operator anticommutes with the supercharges.
Finally, using the supercharges defined above, one may define Hermitian supercharges $Q_H^i$ as
\be Q_H^i=\frac{1}{\sqrt{2}}(Q^i+Q^{i\dag}). \ee

In QM, the algebraic method amounts to the consideration of creation and annihilation operators. For a (one-dimensional) bosonic set of states the relevant operators are
$a=x+ip$, $a^{\dag}=x-ip$, which satisfy the commutation relations 
$[a,a^{\dag}]=1$, $[a,a]=[a^{\dag},a^{\dag}]=0$. 
Similarly, a fermionic set of states may be described using operators $\a,\a^{\dag}$ satisfying the anticommutation relations
$\{\a,\a^{\dag}\}=1$, $\{\a,\a\}=\{\a^{\dag},\a^{\dag}\}=0$.

According to the above, a supersymmetric QM system in one dimension may be associated with two sets 
of operators, $a,a^{\dag}$ and $\a,\a^{\dag}$ with the above properties. The Hamiltonian of the combined system is given by
\be H=H_b+H_f=\hat n_a+\hat n_{\a}, \ee 
where $\hat n_a=a^{\dag}a$ and $\hat n_{\a}=\a^{\dag}\a$ are number operators, 
and it acts on a Hilbert space of quantum states of the form
\be 
|b\rangle\otimes|f\rangle \in {\cal H}={\cal H}_b\oplus {\cal H}_f. 
\ee
The supercharges may in turn be defined as
\be Q=\sqrt 2a^{\dag}\a, \qquad Q^{\dag}=\sqrt 2a\a^{\dag}, \ee
while the Hermitian one takes the form
\be Q_H=a^{\dag}\a+a\a^{\dag}. \ee 
Then it is rather straightforward to show that the square of the Hermitian supercharge is equal to the Hamiltonian of the system,
\be\label{hermscharge} Q_H^2=H. \ee
The simplest quantum mechanical system exhibiting the above algebraic structure is
that of a charged particle moving on a two-dimensional plane under
the influence of a constant magnetic field $\mathbf{B} = B\hat{z}$
perpendicular to it. In that case the Hamiltonian of the system 
is given by
\be
H = H_b + H_f = \hbar\omega_B(a^{\dag}a+\a^{\dag}\a)
\ee
where the bosonic part is given by
\be H_b=\frac{\pi_x^2+\p_y^2}{2m} = \hbar\omega_B(a^{\dag}a+\frac 12), 
\ee
where $\pi_i, i=x,y$ are the canonical momenta, while the fermionic part is 
\be 
H_f=-\mathbf{\mu}\cdot\mathbf{B} = \hbar\omega_B(\a^{\dag}\a-\frac 12), 
\ee
reflecting the coupling of the particle's magnetic moment to the magnetic field.

\paragraph{Supersymmetry and the quantum planes.}

In our framework we do not deal with QM but rather with quantum spacetime, with commutation relations of the form 
\be\label{qplaneappendix} [X^a,X^b]=i\theta^{ab}.
\ee 
Therefore, the operators $a, a^{\dag}$ etc. in our case are not combinations of positions and momenta; 
they strictly combine positions, as  in (\ref{y-oscillator}). However, although the physical interpretation is different, the mathematical similarity may be exploited further. 

Consider the two-dimensional Moyal-Weyl quantum plane $\R_{\theta}^2$ given by (\ref{qplaneappendix}) with $a,b=1,2$. Then in complete analogy with the previous paragraph, let us define the operators
\be a=X^1+iX^2, \qquad a^{\dag}=X^1-iX^2, \ee
as well as the anticommuting operators $\a,\a^{\dag}$  as above. 
These operators satisfy all the properties of the previous paragraph. Moreover, they may be used to define the Hamiltonian $H$, as well as the supercharges $Q,Q^{\dag},Q_H$ exactly as before. Therefore, considering a Hilbert space of bosonic and fermionic quantum states on the $\R_{\theta}^2$, we end up with a system which is formally the same as the supersymmetric QM ones.

Going one step further, let us consider again the two perpendicular quantum planes $\R^2_{45}\cap \R^2_{67}$, as in section 3. 
There we introduced two sets of bosonic oscillators for the ''bosons``, namely $a,a^{\dag},b,b^{\dag}$ and two fermionic ones for the ''fermions``, namely $\a,\a^{\dag},\b,\b^{\dag}$. Let us now show how to realize the supersymmetry algebra (\ref{susyalg}). The supercharges are given by the following operators,
\bea 
	Q_{(1)}&=&\sqrt 2\a a^{\dag}, \qquad Q_{(1)}^{\dag} =\sqrt 2 \a^{\dag}a, \\
	Q_{(2)}&=&\sqrt 2\b b^{\dag}, \qquad Q_{(2)}^{\dag} = \sqrt 2\b^{\dag}b. 
 \eea 
Defining 
\bea Q&=& Q_{(1)}+Q_{(2)}, \\
	Q^{\dag}&=&Q_{(1)}^{\dag}+Q_{(2)}^{\dag},
\eea
it is easy to check that the above operators indeed satisfy the properties of the algebra (\ref{susyalg}) 
for $N=1$,  with Hamiltonian
\be 
H=\hat n_a+\hat n_b+\hat n_{\a}+\hat n_{\b}.
\ee
Let us note that $H$ indeed deserves to be called the Hamiltonian of the system for one more reason apart from being the sum of the 
occupation numbers of a system of two fermionic and two bosonic oscillators. In (\ref{D2-stretch-EV}), 
we observe that 
\be H=\slashed D^2, 
\ee 
as expected.
Therefore the Hamiltonian is equal to the square of the Dirac operator on the quantum planes. This will be very important in the following discussion concerning the fermionic zero-modes.

\paragraph{Fermionic zero-modes.}

Let us finally discuss the relation between the supersymmetric formulation of the previous paragraphs and the existence of fermionic zero-modes on the intersecting solutions of the matrix model. 

In the above discussion of supersymmetric QM, the operator $(-1)^F$ was introduced. The importance of this operator lies in the fact that its trace is in fact topologically stable and it corresponds to the mathematical concept of the index of an elliptic operator. Indeed, as it was shown in \cite{Witten:1982df}, the following relations hold,
\be Tr(-1)^F=n_{b}^0-n_{f}^0=\mbox{index}(\slashed{D}), \ee
where $n_b^0$ and $n_f^0$ denote the zero-energy bosonic and fermionic modes respectively, while $\slashed{D}$ is the Dirac 
operator of the system, 
which is in fact equal to the Hermitian supercharge, i.e. 
\be \slashed{D}=Q_H. \ee
The most important property of the operator $Tr(-1)^F$ is that when it is not zero, supersymmetry is not 
spontaneously broken.

The relevance of the above to our discussion comes along with the formulational coincidence between supersymmetric 
QM and the oscillator representation for the quantum planes. Indeed, in the case of two intersecting quantum planes 
studied in section 3.2.1, we have determined the Dirac operator for the off-diagonal spinors. 
It is given by eq. (\ref{Dirac-action-offdiag-general}). Moreover, in eq. (\ref{D2-stretch-EV}) appears 
the square of the Dirac operator. In the present formulation, these equations may be rewritten as
\bea
\slashed{D}_6\Psi_{(ab)}&=& \big(Q_{H(1)}-Q_{H(2)}\big)\Psi_{(ab)} \nn \\
			&=& \frac 1{\sqrt{2}}\big(Q_{(1)}+Q_{(1)}^{\dag}\big)\Psi_{(ab)}-\frac 1{\sqrt{2}}\big(Q_{(2)}+Q_{(2)}^{\dag}\big)\Psi_{(ab)}
\eea 
and 
\be \slashed{D}_6^2\Psi_{(ab)}=\frac 12 \big(Q_{(1)}+Q_{(1)}^{\dag}\big)^2\Psi_{(ab)}
 +\frac 12 \big(Q_{(2)}+Q_{(2)}^{\dag}\big)^2\Psi_{(ab)}.
\ee
Therefore the presence of zero-modes is equivalent to the conditions 
\bea \label{zeromodesusycondition1}
Q_{H(1)}\Psi_{(ab)}&=&0,\\
\label{zeromodesusycondition2}
Q_{H(2)}\Psi_{(ab)}&=&0,
\eea
which directly imply the result appearing in eq. \eq{localized-state}. In that case, we observe that 
supersymmetry is not spontaneously broken. Indeed, in the present case it holds that $n_b^0=0$, while $n_f^0\ne 0$ and 
therefore $Tr(-1)^F\ne 0$. An alternative way to see this is through the vacuum energy of the system. 
As long as the conditions \eq{zeromodesusycondition1} and \eq{zeromodesusycondition2} hold and since the Hermitian 
supercharge is related to the Hamiltonian of the system as in \eq{hermscharge}, it follows that the vacuum energy vanishes. 
This directly implies the presence of supersymmetry, the vacuum energy being the order parameter for (global) supersymmetry breaking.
Of course, this  quantum mechanical supersymmetry has nothing to do with the space-time supersymmetry 
on $\R^4$.

Evidently, the above discussion can be straightforwardly generalized for any other system of intersecting quantum planes, 
i.e. for more than two quantum planes or for higher-dimensional ones. 
The results are then qualitatively the same as the ones presented here.

\end{document}